\documentclass[10pt]{article}

\usepackage[latin1]{inputenc}
\usepackage[T1]{fontenc}
\usepackage[english]{babel}
\usepackage{amsfonts}
\usepackage{amssymb}
\usepackage{amsmath}
\usepackage{amsthm}
\usepackage{graphicx}
\usepackage{fullpage}

\def\be{\begin{equation}}
\def\ee{\end{equation}}
\def\bc{\begin{center}}
\def\ec{\end{center}}

\title{A thermodynamical perspective of immune capabilities}

\author{Elena Agliari\footnote{Dipartimento di Fisica,  Universit\`a di Parma, viale G.P. Usberti 7/A, 43100 Parma (Italy) and INFN, Gruppo collegato di Parma (Italy)}, Adriano Barra\footnote{Dipartimento di Fisica, Sapienza Universit\`a di Roma, Piazzale A. Moro 2, 00185 Roma (Italy) and  GNFM, Gruppo di Roma1 (Italy)}  \ Francesco Guerra\footnote{Dipartimento di Fisica, Sapienza Universit\`a di Roma, Piazzale A. Moro 2, 00185 Roma (Italy) and INFN, Gruppo di
Roma1 (Italy)}, \ Francesco Moauro\footnote{Dipartimento di Fisica, Sapienza Universit\`a di Roma, Piazzale A. Moro 2, 00185 Roma (Italy)}}

\begin{document}

\maketitle

\begin{abstract}
We consider the mutual interactions, via cytokine exchanges, among helper lymphocytes, B lymphocytes and killer lymphocytes, and we model them as a unique system by means of a tripartite network. Each part includes all the  different clones of the same lymphatic subpopulation, whose couplings to the others are either excitatory or inhibitory (mirroring elicitation and suppression by cytokine). First of all, we show that this system can be mapped into an associative neural network, where helper cells directly interact with each other and are able to secrete cytokines according to ``strategies'' learnt by the system and profitable to cope with possible antigenic stimulation; the ability of such a retrieval corresponds to a healthy reaction of the immune system. We then investigate the possible conditions for the failure of a correct retrieval and distinguish between the following outcomes: massive lymphocyte expansion/suppression (e.g. lymphoproliferative syndromes), subpopulation unbalance (e.g. HIV, EBV infections) and ageing (thought of as noise growth); the correlation of such states to auto-immune diseases is also highlighted.
Lastly, we discuss how self-regulatory effects within each effector branch (i.e. B and killer lymphocytes) can be modeled in terms of a stochastic process, ultimately providing a consistent bridge between the tripartite-network approach introduced here and the immune networks developed in the last decades.\\
Keywords: Theoretical Immunology, Statistical Mechanics, Autoimmunity
\end{abstract}

\section{Introduction}
The immune system is one of the most advanced and complex biological systems, made up of many different kinds of cells, and hundreds of different chemical messengers, which must be properly orchestrated for ensuring a safe collective performance, that is, to protect the host body against foreign organisms and substances, also recognizing objects as either damaging or non-damaging. The system includes  different classes of cells working as ``soldiers'' and different classes of proteins working as ``weapons'', each carrying out specialized functions (e.g. alert, activate, engulf, kill, clean up, etc.); all the immune cells synthesize and secrete special proteins that act as antibodies, regulators, helpers or suppressors of other cells in the whole process of defending against invaders.

Like the nervous system, the immune system performs pattern recognition, learns and retains a memory of the antigens that it has fought off. Accomplishing such complex tasks requires the cooperation (via cell-to-cell contacts and exchanges of secreted messengers) among a huge number of components and this allows for using the methods and the concepts of statistical mechanics.
Indeed, a systemic viewpoint, embedded in a statistical mechanics framework, may be a strategic approach to evidence which are the key mechanisms underlining the (mis)functioning of the system and therefore to prevent  diseases and derangements.

Basically,  the architecture of the model we introduce keeps track of the (manifest) interactions among agents, while statistical mechanics gives the rules, through thermodynamical variational principles, and ultimately allows to uncover the key mechanism and, possibly, hidden correlations.

Beyond a general picture, the particular phenomenon we want to deepen and explain in terms of a cooperative behavior of immune cells is the emergence and the effects of lymphocytosis (i.e. an abnormal immune response by lymphocytes): The correlation between a strong lymphocytosis and autoimmune manifestations is a well-known experimental finding for which a plethora of interpretations and descriptions have been provided, yet a unifying, sytematic picture is still missing \cite{abbas,a3,a6,a8,b1,b2,b3}.
Essentially, two types of lymphocytosis exist: as a response to a pathogen
(i.e. Epstein-Barr virus, EBV, or Human immunodeficiency virus, HIV, etc), which may affect the host for a while and then disappear, becoming latent (and is coupled with a short term or pulsed autoimmunity) \cite{b4,aids}, and
as Autoimmune Lymphoproliferative Syndrome (ALPS), a chronic lymphocytosis due
to the lack, in killer cells, of the Fas genetic expression, a regulatory intracellular mechanism  that induces the apoptosis \cite{c3,c6}. ALPS is a severe disease which
develops in strong, persistent, autoimmune manifestations and, despite well understood at micro (genetic) and macro (clinical) levels of description, a consistent, merging description is still missing \cite{infants}.

The model we introduce includes effector cells (killer lymphocytes $\mathrm{T_K}$ and B cells) and helper cells $\mathrm{T_H}$, whose mutual interactions, occurring via cytokines exchange, give rise to a network, where helpers are connected to both effector cells (B, $\mathrm{T_K}$), while there is no direct connection among the latter. Since the effect of cytokines exchanged can be either excitatory or inhibitory, this realizes a tripartite spin-glass system. We firstly show that such a system is equivalent to an associative neural network where helper cells are able, thanks to a cooperative synergy, to perform retrieval of  ``strategies'' learnt by the system and profitable to cope with possible antigenic stimulation.  Hence, the ability of such a retrieval corresponds to a healthy reaction of the immune system.

A state of poly-clonal  lymphocytosis is then realized by increasing the average size of effector populations, hence mimicking a persistent clonal expansion.  Interestingly, we find that this alteration is formally equivalent to a random field acting on helper cells, which induces ``disorder'' within the system. As we will explain, this can be read from a thermodynamic perspective: The presence of a (sufficiently large) antigenic concentration induces the system to do some ``work'' (a clonal expansion), which turns out to be split in an internal energy term (necessary to make the network able to
recognize) and in a heat term (emerging as an unavoidably feature of this conversion). Otherwise stated,
it is not possible to obtain an extensive immune response (a clonal expansion of B or $\mathrm{T_K}$
cells), which play the role of a ``work'' (as it is an ordered result), without introducing
some noise (heat) in the network of interacting cells, the whole resembling the well known principles  of thermodynamics.

Furthermore, we find that the average extent of effector population must range within a given interval in order for the system to be performing: If the population is too small, the interplay among components is too weak in order to establish a mutual interaction and regulation; on the contrary, if, e.g. due to a strong immune response, the population gets too large, the level
of noise in the network may become so high
that the system starts to fail to select the right strategies to fight, and ultimately, it
attacks the self, producing the autoimmune response. Similarly, we will show that if the balance between lymphocyte sub-populations is lost (as it happens in HIV and EBV infections), another kind of ``noise'' prevails and, again, the system is no longer able to work correctly.
Bad signaling can also prevail due to a progressive growth of the randomness in the stochastic process we consider; increasing white noise corresponds to an aging process, consistently with the evidenced malfunctioning of elder systems: Debris from e.g. lysis of infected cells by killers, may act as ``dust'' in the `gears'' of pattern recognitions.

Hence, our model highlights three ways to escape from the ``healthy region'': massive clonal expansion/suppression, unbalance in subpopulations, aging.
Despite not yet quantitatively comparable to real data, a clear theory of this mechanism opens completely new paths to deal with autoimmune
diseases, which affects almost
one fourth of the worldwide population \cite{AI}.

The paper is organized as follows:  in Sec. 2 we describe, from an immunological point of view, the agents making up the system we are focusing on; in Sec. 3 we describe in details the formal model used to describe the system itself, and in Sec. 4 we analyze its behavior, stressing the conditions leading to an incorrect performance; our conclusions and perspectives are collected in Sec. 5. The technical passages involved in the statistical mechanics analysis of our model are gathered in the appendices, together with a discussion on the hidden effects of mutual interactions within the effector branches.

\section{The immunological scenario} \label{sec:immuno}
The system we are focusing on provides a modeling for the interplay among lymphocytes (B cells, $\mathrm{T_K}$ cells and $\mathrm{T_H}$ cells) mediated by cytokines (interleukin family, interferon family, etc.); before proceeding, we sketch the main functions of such agents \cite{abbas} and of some pathologies (e.g. HIV infection or ALPS), which stem from an improper functioning of these agents.

\bigskip

\emph{B lymphocytes.}
The main role of these cells is to make antibodies (primarily against antigens \footnote{A healthy immune system produces also a small amount of self-reactive lymphocytes, whose antibody production is low and regulated by the network of cells making up the whole system, so that it is not dangerous for the host body \cite{abbas}.}) and to develop into memory B cells after activation by antigen interaction.
When receptors on the surface of a B cell match the antigens present in the body, the B cell (aided by helper T cells) proliferates and differentiates into effector cells, which secrete antibodies with binding sites identical to those displayed by receptors on the ancestor-cell surface (hypersomatic mutation apart \cite{abbas}), and into memory cells, which survive for years preserving the ability to recognize the same antigen during a possible re-exposure. According to the shape of the receptors they display, B cells are divided into clones: cells belonging to the same clone can recognize and bind the same specific  macromolecules (epitopes) of a given antigen.
A set of up to $10^9$ different clones allows for deeply diverse and specific immune responses.

Another important role of B cells is to perform as antigen-presenting cells (APC) to other agents; this makes B cells able to interact with (mature) T cells, through the so-called ``immunological synapse''.
Indeed, the recognition of an antigen is not sufficient for B cell activation: An additional signal from helper T cells is in order and this is realized by means of chemical messengers (cytochines, see below) secreted by the matching T cell \cite{giorgio}.

\bigskip

\emph{$\mathrm{T_K}$ lymphocytes.}
Cytotoxic CD8+ cells (also known as ``Killer cells") belong to the group of T lymphocytes. $\mathrm{T_K}$ cells are capable of inducing the death of infected, tumoral, damaged or dysfunctional cells. Analogously to B cells, the activation of cytotoxic T cells requires not only the presence of the antigen, but also a second signal provided by the cytokines released from helper T cells.

 In fact, cytotoxic T cells express receptors (TCRs) that can recognize a specific antigenic peptide bound to the so called class-I MHC molecules (present on nearly every cell of the body); upon recognition CD8+ cells are regulated by the chemical messengers (cytochines, see below) secreted by active helper T cells.
More precisely, CD8+ cells undergo clonal expansion and differentiation into memory and effector cells with the help of a cytokine called Interleukin-$2$; as a result, the number of effector cells for the target antigen increases and they can then travel throughout the body in search of antigen-positive cells.

\bigskip

\emph{$\mathrm{T_H}$ lymphocytes.}
Helper CD4+ cells  are a sub-group of T lymphocytes that play a crucial role in optimizing the performance of the immune system. These cells do not posses cytotoxic or phagocytic activity, neither they can  produce antibodies, yet, they are actually fundamental for regulation of the effector branches of the immunity and this job is basically accomplished by secretion and absorbtion of cytokines.

CD4+ T cells exhibit TCRs with high affinity for the so called class-II MHC proteins, generally found on the surface of specialized APCs, e.g. dendritic cells, macrophages and B cells. The presentation of antigenic peptides from APCs to CD4+ T cells provides the first signal, which ensures that only a T cell with a TCR specific to that peptide is activated.
The second signal involves an interaction between specific surface receptors on the CD4+ and on APC and it licenses the T cell to respond to an antigen. Without it, the T cell becomes anergic, and does not respond to any antigen stimulation, even if both signals are present later on. This mechanism prevents inappropriate responses to self, as self-peptides are not usually presented with suitable co-stimulations \cite{kitamura}.

Once the two signal activation is complete, the T helper cell  proliferates and  releases and/or absorbs regulatory agents called cytokines, then they differentiate into the subfamilies $\textrm{T}_{\textrm{H}_1}$ or $\textrm{T}_{\textrm{H}_2}$ depending on cytokine environment (however for our purposes this further distinction is not needed).

\bigskip

\emph{Cytokines.}
Cytokines are small cell-signaling proteins secreted and absorbed by numerous cells of the immune system, functioning as intercellular messengers: Cytokines include e.g. interleukins, interferons and chemochines.

These are usually produced by stimulated cells and are able to modify the behavior of secreting cells themselves (autocrine effect) or of others (paracrine effect), not necessarily spatially  close (endocrine effect), inducing growth, differentiation or death. Most cytokines are produced by CD4+ cells and to a lesser degree by monocytes and macrophages.
In general, cytokines display both agonist and antagonist action in such a way that if the pro-inflammatory milieu exceeds that of the anti-inflammatory mediators the net effect will result in prolonged inflammation. 

Moreover, cytokines attach to receptors on the outside of cells causing the target cell to produce other cytokines;  this complicated relationship is called the cytokine network, and it is one of the most important ways used by the immune system, spread throughout the body, to communicate and orchestrate appropriate responses to the various challenges. Indeed, cytokines act as key communicators for immune cells and the delicate balance in the level of these communicators is vital for health: Many chronic diseases arise due to a disruption of this balance \cite{cytokines}.

In fact, plasma levels of various cytokines may give information on the presence of inflammatory processes involved in autoimmune diseases such as rheumatoid arthritis, as well as immunomodulatory effects of foods or drugs; in addition,  cytokines play an important part in the progression from HIV infection to AIDS disease and in many AIDS-related illnesses: In particular, the initial HIV infection disrupts the normal balance of cytokines by causing the levels of certain cytokines to rise; cytokine imbalances then helps HIV to target CD4+  cells and the lymph nodes, leading to the progressive immunosuppression and the opportunistic infections that follow \cite{aids,a4}.

\bigskip

\emph{Autoimmune diseases.}
These pathologies stem from a failure of the immune system to recognize its host body as self, and this causes an immune response against host's own cells and tissues \cite{a3,a6,a8,b1,b2,b3}. In the early development of theoretical immunology the main strand to tackle the problem of self/non-self discrimination by immune cells was the clonal deletion: In the bone-marrow (for the B) or in the thymus (for the T) all the cells are tested and those self-reacting are deleted in such a way that cells making up the effective repertoire are only those specific for foreign antigens. Actually, it is now established (both experimentally and theoretically) that a low level of self-reactivity is normal and even necessary for the immune system to work properly. From a systemic point of view it is just the synergic interplay between immune agents, in particular cytokines, immunoglobulins and lymphocytes, that keeps the concentrations of self-reactive cells at the right values \cite{a8,b2,noi1,noi2}.

\bigskip

\emph{Lymphocytosis.}
Lymphocytosis refers to an abnormal increase in the count of white cells (i.e. $\geq 10^9$ cells/litre \cite{abbas}), sometimes evolving in chronic lymphocytic leukemia. Lymphocytosis can be essentially
of two types, either mono-clonal or poly-clonal, referring to the expansion
of a particular clone or of an ensemble; in this work we deal with the latter.
This pathology essentially happens as a response to particularly ``smart'' antigens (i.e. EBV, HIV, etc), as a chronic condition of long term infections (i.e. tuberculosis, brucellosis, syphilis), from malignancies (i.e. leukemia) or, lastly, due to a genetic alteration causing the lack of lymphocyte apoptosis after their clonal expansion (autoimmune lymphoproliferative syndrome, ALPS). More precisely, ALPS patients have been found to carry mutations in genes Fas and FasL, which are upstream effectors of the apoptotic pathway \cite{c3,c6}.
This inefficacy of apoptosis causes an increase in the number of lymphocytes in the body, including cells that are too old and less effective, and a consequential bad regulation of cytokines secretion \cite{a1,a2,a5}. As a result, there is the failure of immunological homeostasis, possibly leading to autoimmunity, and the development of lymphoma.
ALPS can be diagnosed by blood tests; it occurs in both sexes and has been described in patients (mostly children) from all over the world. ALPS is a rare condition which has been defined only within the past few years; its incidence has not yet been estimated \cite{jack,rieux}.

\section{The model} \label{sec:model}
As anticipated, the protagonists of our model are B cells (which produce antibodies), CD8+ cells (``killers'' which delete infected cells), and CD4+ cells (``helpers'' which coordinate the two effector branches) of the immune response. Each type is constituted by clones with a given specificity and the overall number of different clones is denoted as $B$, $K$ and $H$, respectively. The number of cells making up a given clone is not constant in time, but may increase due to an antigenic stimulation addressed to the pertaining specificity (the Burnet clonal expansion \cite{burnet}); in the following we call ``activity'' a logarithmic measure of the amplitude of a clone and we denote it by the set of variables $b_{\nu}$ ($\nu \in (1,...,B)$), $k_{\mu}$ ($\mu \in (1,...,K)$), and $h_i$ ($i \in (1,...,H)$), in such a way that the actual concentration of B cells, killers and helpers is $\sim \sum_{\nu} \exp(b_{\nu}), \sim \sum_{\mu} \exp(k_{\mu}), \sim \sum_{i} \exp(h_i)$, respectively \cite{noi1,noi2,aldo}.

\begin{figure}
 \begin{center}
\includegraphics[width=.48\textwidth]{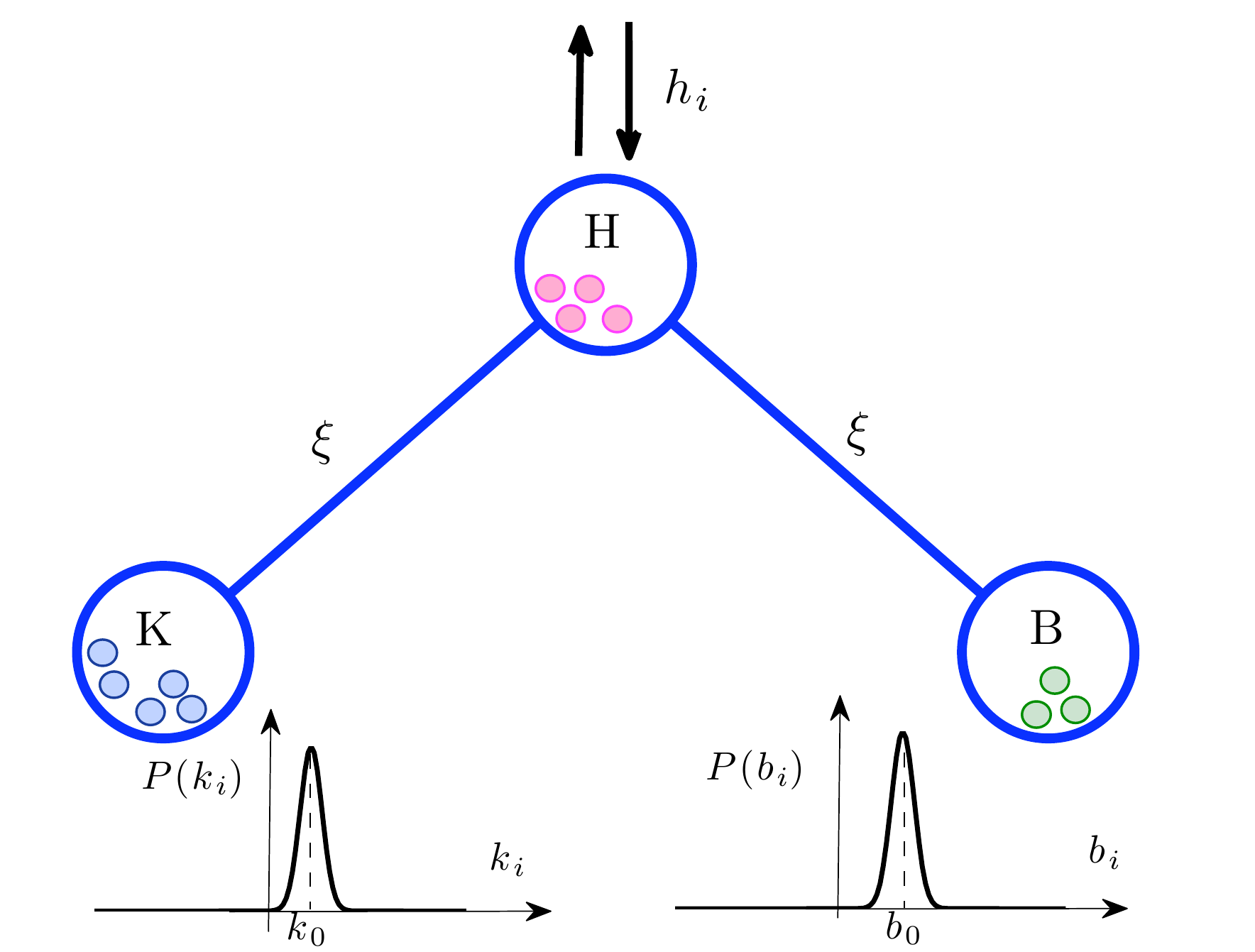}
\caption{\label{fig:tri} (Color on line) Schematic representation of the tripartite system considered. Each node envisages a set of $B, H, K$ different clones of B, CD4+ and CD8+ cells respectively. The activity of each clone is described by a dichotomic variable $h_i$ or by a Gaussian variable $b_i$ and $k_i$ centered around a value $k_0$ and $b_0$, respectively.}
\end{center}
\end{figure}

Of course, different clones of the same branch can interact with each other (see e.g. \cite{noi1,noi2,jerne,antonio,parisi}); these interactions, at least for B and $\mathrm{T_K}$, play a role in the development of memory of previous antigens or in self/non-self discrimination and it is effectively accounted for by taking Gaussian distributions for their activity (as explained in Appendix A), yet, here we focus on the interactions mediated by cytokines which provide signals acting between CD4+ and B cells, as well as between CD4+ and CD8+ cells.
Such interactions give rise to a tripartite network, where parties are made up of CD4+, B and CD8+ clones, respectively, and links are drawn whenever interleukins and/or interferons are exchanged among them (see Fig.~\ref{fig:tri}).

Now, as anticipated, the activity of both effector branches (i.e. CD8+ and B cells), is
assumed to be distributed around a given mean value, which, at equilibrium, must be very small denoting a typical low activity; in agreement with experimental findings \cite{jensen} and with the arguments developed in Appendix A, we say that, at rest, (for any $\mu,\nu$) $k_{\mu}$ and $b_{\nu}$ follow a Gaussian distribution $\mathcal{N}[0, 1]$, peaked at zero and with unitary standard deviation. On the other hand, CD4+ cells are described by dichotomic variables, that is $h_i = \pm 1$, for any $i$\footnote{This assumption implicitly suggests that the time scale for helper reaction is slower than those of the branches: indeed in our perspective the latter act as inputs of information for helpers which then need further time to elaborate and readjust their state; this is also consistent with the fact that, due to the interplay of the subfamilies $T_{H_1},T_{H_2}$, the time needed by helper cells to respond to a stimulation is relatively large (see for instance the discussion in \cite{pavlov}). Furthermore, if the timescale for helper cells was faster, their contribution to the potentials $V(\mathbf{b})$ and $V(\mathbf{k})$, introduced in appendix A and ruling the activity of effector cells, would be zero since the integral over helper states is null.}; positive values mean that the relevant clone is in an active state, namely high rate of cytokines production, viceversa $-1$ stands for quiescence.
Anyhow, it is worth underlining that the picture we are going to offer does not depend qualitatively on the kind of distribution, either Gaussian or binary, chosen for the activity\footnote{Here we choose the latter for helpers as
the discrete nature of $h$ variables is consistent with integrate-and-fire models \cite{amit}, where the action of the agent considered (e.g. a neuron, a lymphocyte, etc.) is generated when the received input (e.g. a voltage, a cytokine concentration, etc.) reaches a threshold, as this will simplify the mathematical handling of our model later on.}.

Finally, the state of cytokines acting between the $i$-th helper clone and the $\mu$-th $B$ or $\nu$-th killer clone, is encoded by a dichotomic variable $\xi_{i \mu} = \pm 1, \xi_{i \nu} = \pm 1$; a positive (negative) value means that there is an excitatory (inhibitory) stimulation. Here we adopt a minimal assumption and we say that the probability distributions for $\xi$ is given by $P(\xi_{i,\mu/\nu}=1) = P(\xi_{i,\mu/\nu}=-1) = 1/2$\footnote{The assumption of identically, symmetrically distributed cytokines can be possibly relaxed as done for instance by Amit \cite{amit} in the neural scenario to reflect more realistic biological features.}.
In the following, the cytokine pattern $\{ \xi_{i \mu},  \xi_{i \nu} \}$ is supposed to be \emph{quenched} and, as we will see, such pattern encodes proper ``strategies'' learnt by the system during life and profitable to cope with possible antigenic stimulation\footnote{So far we assumed that learning is already achieved during ontogenesis and we focus only on the ability in retrieval of mature immune systems. However, the model we present, whose associative memory ability is obtained bridging it to an associative neural network, is a tripartite spin-glass that naturally describes a two-layered restricted Boltzmann  machine as those commonly used to store information  in Machine Learning \cite{machinelearning,benjo}.}.

\subsection{The statistical mechanics approach}
The following analysis addresses the functioning of the system considered, that is, we look at the conditions for the establishment of stable configurations and/or for the rearrangement of configurations following an external stimulation.

In order to describe the system we introduce the ``Hamiltonian''
 \begin{equation} \label{eq:H}
 \mathcal{H}_{H,B,K}(h,k,b;\xi) =
\frac{-1}{\sqrt{H}}\sum_{i,\mu}^{H,K}\xi_{i,\mu}h_i k_{\mu} - \frac{1}{\sqrt{H}}\sum_{i,\nu}^{H,B}\xi_{i,\nu}h_i b_{\nu},
\end{equation}
where the first term accounts for the interactions between clones of CD4+ and CD8+ populations, while the second term for the interactions between clones of CD4+ and B populations.
Following statistical mechanics prescriptions, the Hamiltonian $\mathcal{H}_{H,B,K}(h,k,b;\xi)$\footnote{The system considered admits an Hamiltonian representation due to the assumed symmetry in couplings which ensures detailed balance, which, in turn, ensures the existence of canonical equilibrium. Surely a step forward would be the off-equilibrium analysis of the non-symmetrical version of the model, which may also allow a quantitative matching with experimental data.} is nothing but a cost function for the configuration $\{h_i, k_{\mu}, b_{\nu} \}$: The smaller its value and the more likely the correspondent configuration; in the jargon of disordered system, Eq. (\ref{eq:H}) represents a tripartite spin-glass.

Since the statistical mechanics analysis will be performed in the infinite system size limit, i.e. in the limit of large $H, B$ and $K$, we need to specify a meaningful scaling for their ratios by introducing the parameters $\alpha, \gamma \in \mathbf{R}^+$ such that
\begin{equation}\label{eg:hl}
\alpha \equiv \lim_{H \rightarrow \infty} \frac{B}{H}, \; \; \; \gamma \equiv \lim_{H \rightarrow \infty} \frac{K}{H}.
\end{equation}

The performance of the system described by Eq. (\ref{eq:H}) can then be studied following the standard routine used for disordered systems \cite{amit,peter}. First of all, one calculates the partition function
\begin{eqnarray}\label{partition}
Z_{H,B,K}(\beta; \xi) &=& \sum_{\{h\}} \int \prod_{\nu=1}^B d\mu(b_{\nu})\int
\prod_{\mu=1}^K d\mu(k_{\mu})
\exp\left[ -\beta \mathcal{H}_{H,B,K}(h,b,k;\xi) \right],
\end{eqnarray}
where $d \mu(x)$ represents the Gaussian measure on $x$, that is $\exp(- x^2/2)$, and $\beta \in \mathbf{R}^+$ is the degree of (white) noise in the network: Large values of $\beta$ (small noise limit) make the Boltzmann weight $\exp(-\beta \mathcal{H})$ more significant.
From the partition function all the thermodynamic observables can be derived: for the generic continuous function  $O(h,b,k)$ one has the so-called Boltzmann state $\omega (O)$ as
\begin{equation}
\omega(O)=\frac{1}{Z_{H,B,K}(\beta;\xi)} \sum_{\{h\}} \int \prod_{\nu=1}^B d \mu(b_{\nu}) \int \prod_{\mu=1}^K d \nu(k_{\mu}) O(h,b,k) \exp \left[ - \beta \mathcal{H}_{H,B,K}(h,b,k;\xi) \right].
\end{equation}
For our concerns, the main quantity of interest is the free-energy (or pressure), which here, in the infinite system size limit, can be evaluated as
\be A_{\alpha, \gamma}(\beta) = \lim_{H \rightarrow \infty}  \frac1H
\mathbb{E}\log Z_{H,B,K}(\beta;\xi), \ee
where we also applied the average $\mathbb{E}$ over all the (quenched) values of $\{ \xi_{i \mu}, \xi_{i \nu} \}$, in order to get an estimate on the typical realization of the cytokine network.
\newline
Now, the minimization of the free energy with respect to proper order parameters (see appendix B) is the
main path to follow in order to get the most likely states, as this contemporary entails energy minimization and entropy maximization; moreover, singularities in any derivative of the free-energy are signatures of phase transitions demarcating regions (in the $\alpha, \beta, \gamma$ space) where the system may or may not work cooperatively (see e.g. \cite{amit}).

Before analyzing $A_{\alpha, \gamma}(\beta)$, we need to compute explicitly the partition function $Z_{H,B,K}$; by noticing that it  does not
involve two-body interactions within each $B,T_K$ branch, but only one-body terms in $b$ and $k$, we can directly carry out the relevant Gaussian integrals so to get
$Z_{H,B,K} = \sum_{\{ h \}} \exp (- \mathcal{H}_{\textrm{Hopfield}}(h; \xi))$,
being $\mathcal{H}_{\textrm{Hopfield}}$ the effective Hamiltonian
\be \label{eq:2hop} \mathcal{H}_{\textrm{Hopfield}}(h;\xi)= -\frac{\beta}{H}
\sum_{i<j}^H \Big( \sum_{\mu=1}^K \xi_i^{\mu}\xi_j^{\mu} +
\sum_{\nu=1}^B \xi_i^{\nu}\xi_j^{\nu}\Big)h_i h_j, \ee
which, interestingly, recovers the Hopfield representation of a neural network (see next subsection) \cite{amit,hopfield}.

Such equivalence states that a $n$-partite spin-glass and a sum of $n-1$ independent neural networks display equivalent thermodynamic behaviors (same free-energy, phase diagram, collective properties, etc.).
\newline
From an immunological perspective, we see that the behavior of a system where
helpers promote/suppress, via cytokines, the two effector branches underlies
an effective system where cytokines directly connect helper cells via an Hebbian interaction making them able to learn, store and retrieve patterns of branch activations: B and $\mathrm{T_K}$ branches work as sources of information (stimulative layers in the neuronal counterpart \cite{machinelearning}) for $\mathrm{T_H}$'s, which, in turn, store such information trough effective pairwise interactions (similarly to the mechanisms applied by neurons \cite{amit}).

Finally, we would like to emphasize that this model, although being a minimal one, is able to capture the crucial traits of the system: Qualitative differences in time-scales as well as the basic properties of interactions are accounted for, also showing that the description obtained is robust with respect to some technical details. Moreover, as we will show, the model displays several \emph{emerging} properties, finely matching real systems.
By the way, the same tripartite system proved to be a proper effective model also for explaining the emergence of Chronic Fatigue Syndrome \cite{pavlov}: In that case the focus was on the evolution of the synapse realized by CD4+ cells between CD8+ and B cells, and on the possibility to develop a Pavlovian associative learning of a prolonged infected status.

\subsection{On the mapping with the neural network}
As already underlined in the past (see e.g. \cite{noi1, parisi,pavlov,hoffmann, net1, net2,net3}), there is a strong analogy between neural and immune systems: Both are able to learn from previous experiences and to exhibit features of associative memory as pattern recognition \cite{noi1,amit,peter,a11}.  In the following we briefly sketch how neural networks are formalized and how they do perform, also clarifying step by step the immunological counterpart within our model.

Hopfield neural networks consist of interacting neurons described by the Hamiltonian (see e.g. \cite{amit,a14,a15})
$$\mathcal{H}_N= - \frac1N \sum_{i<j} J_{ij} \sigma_i \sigma_j,$$
where the states ($\sigma_i = \pm 1$)  represent the two main levels of activity (i.e. firing/not-firing a spike) of the corresponding neuron, while the coupling $J_{ij}$ are the synaptic couplings between pairs of neurons $(i, j)$. Moreover, one considers $P$ ``patterns'' (denoted as $\{ \xi_i^{\mu} \}_{\mu=1,2,...,P}$), which represent the embedded memorizable information and assumes that, as a result of a learning process, the synapses $J_{ij}(\{ \xi \})$ bear values which ensure the dynamic stability of certain neuronal configurations $\{ \sigma_i \}$, corresponding to the memorized patterns; in this sense the network displays associative memory. In the standard theoretical analysis, the $P$ patterns are usually taken quenched\footnote{It is usually assumed that the performance of the network can be analyzed keeping the synaptic values fixed, or quenched. This implies that during a typical retrieval time the changes that may occur in synaptic values are negligible.} and random, with equal probabilities for $\xi_i^{\mu} = \pm 1$.
The specific form of storage prescription usually considered is given by the Hebbian learning rule
\begin{equation}\label{eq:heb}
J_{ij} = \sum_{\mu=1}^P \xi_i^{\mu} \xi_j^{\mu},
\end{equation}
and it is straightforward to see that, by plugging Eq. (\ref{eq:heb}) into $\mathcal{H}_N$, once having renamed $P$ as $B$ and $K$ one recovers each of the terms in Eq. (\ref{eq:2hop}).

Hence, in the immunological scenario, each CD4+ cell plays the role of a neuron, and its state ($h_i = \pm 1$) represents the two main levels of activity (i.e. secreting/not-secreting a cytokine signal) for the corresponding specificity. The coupling $J_{ij}$ derives from the combination of the set of cytokines secreted by clones $i$ and $j$, respectively: if these clones interact in the same way ($\xi_i^{\mu} = \xi_j^{\mu}=\pm1$) with the clone $k_{\mu}$ ($b_{\mu}$) there is a positive contribute to the coupling $J_{ij}$ and vice versa. In this way the interaction between different kinds of lymphocytes is bypassed and allowed for by a direct interaction between CD4+ cells only, which effectively coordinate the effector responses.

Let us now consider memory features: Patterns are said to be memorized when every network configuration $\sigma_i = \xi_i^{\mu}$ for $i=1,...,N$ for every of the $P$ patterns labeled by $\mu$, corresponds to free-energy minima (i.e. stable to all single-state flips), also called attractors.
In the immunological scenario patterns can still be thought of as  the ``background'' of the system, that is, they encode successful ``strategies'' adopted by the system during previous diseases and infections and then properly stored for being retrieved in case of future attack by those antigens.
Therefore, we define a ``strategy'' as a  pattern of information that the helpers send (exchange) to a particular clone of a branch: If, for example, focusing only on B $- \textrm{T}_H$ interactions, the system wants to tackle a response against an antigen and, say, both clones $\nu_1$ and $\nu_2$ are able to bind to it, then helpers are expected to arrange in such a way that, according to the values of $\{ \xi_i^{\nu_1}, \xi_i^{\nu_2}\}$, both $b_{\nu_1}$ and $b_{\nu_2}$ result to be excited, conversely, B cells which do not bind to the antigen are not involved as they receive suppressive signals\footnote{Within our framework, the need for suppression of non-involved clones is clear as they would contribute only raising the noise level, implying bad functioning and dangerous correlations. This is in agreement with the experimental finding that (leukemia or lymphocytosis apart which imply a pathological activation of the immune system) the amount of lymphocytes in the blood is roughly constant over time (ranging from $\mathcal{O}(10^{12})$ to $\mathcal{O}(10^{14})$, that means on logarithmic average  $13 \pm 1$), which means that only a very small number of families is activated.}. Of course, the implicit assumption is that the antigen considered has already been dealt with and the strategy $\{ \xi_i^{\nu_1}, \xi_i^{\nu_2}\}$ has been properly stored.

We recall that the retrieval of a strategy is realized when the corresponding activity configuration for helper lymphocytes, i.e. $\{h_i\}$, is stable in time, which means that CD4+ cells secrete and absorb cytokines in a collective fashion. Now, despite the fact that the $J_{ij}$'s have been constructed to guarantee certain specified patterns to be attractors, namely fixed points of the dynamics, the non-linearity of the dynamical process induces additional non-global minima (linear combinations of pure states), referred to as \emph{spurious states} \cite{amit}: In neural networks such states are considered as erroneous retrieval of an attractor, because the system is meant to retrieve a given pattern of information at each time, while, in the immunological counterpart, their interpretation is rather different as their existence allows for broad, parallel immune responses. More precisely, spurious states realize the overlap of several strategies so that the immune response can address contemporary different kinds of antigen infections. For instance, if two antigens are contemporary present, each with high chemical affinity with, say, three different lymphocytes, the helpers would perform parallel six-strategies by sending the correct signals to the involved cells, eliciting the useful cells and suppressing the non-involved ones.
  Incidentally, we notice that Hopfield networks (as actually close to spin-glasses where the amount of minima scales exponentially with the volume)  work much better as spurious state collectors than pure state retrievers.

In order to quantify the ability of these models in working as associative networks, we consider a set of order parameters (see also next section), among which the $B+K$ Mattis magnetizations $m^{\mu}, m^{\nu}$ which measure the \emph{overlap} of the actual configuration of the helpers with the $\mu$-th and $\nu$-th pattern, that is
\begin{equation}
m^{\mu}  \equiv  \frac{1}{H}  \sum_{i=1}^H \xi_i^{\mu} h_i, \;\;\; m^{\nu}  \equiv  \frac{1}{H}  \sum_{i=1}^H \xi_i^{\nu} h_i,
\end{equation}
both ranging in $[-1,1]$. It is easy to see that if the configuration is correlated (uncorrelated) with a given pattern $\mu,\nu$, the corresponding overlap is macroscopic (vanishes $\sim \mathcal{O}(1/N)$); of course, for full correlation the overlap is unity.

By tuning the parameters of the system, i.e. the noise level $\beta$ and the relative size of the branches with respect to $H$ (namely $\alpha,\gamma$), its ability to retrieve varies significantly. Starting from a high level of noise in a network with a fixed (relative) number of patters $\alpha+\gamma$, there exists an ergodic phase and no retrieval can be accomplished ($m_{\mu,\nu}=0$); indeed, in the limit $\beta \rightarrow 0$, any configuration is equally likely. By decreasing the noise level one crosses a ``spin-glass'' phase\footnote{Spin-glasses are complex systems which, above a certain noise threshold (freezing temperature), are ergodic with spins randomly oriented (paramagnetic phase); at low noise, spin-glasses display a non-ergodic behavior characterized by an enormous amount of metastable configurations, due to the contradictory (frustrating) interactions preventing long range correlations between the orientations of the different spins. Hence, freezing takes place with the spins oriented at random with respect to each other; the magnetization upon freezing is therefore zero, just like in the paramagnetic state. Therefore, in order to evaluate spin-glass emergence, additional order parameters are necessary, often denoted with $q$, such that their average can discriminate between spin-glass freezing ($q>0$) and paramagnetism ($q>0$), (see also next section).}; the noise level at which this happens is $\beta_G(\alpha+\gamma)$. Below this line there is no retrieval ($m=0$), yet the system is no longer full-ergodic. 
Now, if the number of patterns is larger than a certain critical value, i.e. $(\alpha + \gamma)> [\alpha_c(\beta=\infty) + \gamma_c (\beta=\infty)] = 0.138$\footnote{We recall that these values have been calculated for a fully-connected (FC) network, which means than each agent $i$ is connected to any other agent $j \neq i$. Actually, real systems display a non-negligible degree of dilution, for this reason a quantitative comparison should be carried out only once the theory for diluted system will be accomplished, on which we plan to report soon. Here we just mention that values calculated for FC systems underestimate experimental measures and we know that, indeed, the introduction of dilution yields a rise in the critical values \cite{sampo}.}, the reduction of noise is useless for retrieval. While the existence of such a threshold in $\alpha$ is rather intuitive in neural networks (because if  we try to store too many patterns, then the interference among them becomes large, making them ultimately unrecognizable),
here $(\alpha + \gamma)= (B+K)/H$ represents the relative ratio among the inner and the effector branches: As the amount of helpers decreases, the network falls off the retrieval region and the system is no longer able to display a collective performance;  this situation closely resembles the transition from HIV infection to AIDS (we recall that the immunodeficiency virus kills CD4+ cells). Furthermore, we notice a nice consistency between the requirement of a low ratio between the effectors and the helpers expected for a healthy performing system and the fact that antigen recognition is actually spread over the whole lymphocyte network.
In fact, the length of an antibody is $L\sim 10^2$ epitopes; without recognition spreading (i.e. according to a single particle approach) the system would need $\mathcal{O}(2^L)=\mathcal{O}(2^{100})$ different clones to manage antigen attacks. Conversely, the total amount of lymphocytes is estimated to be $\mathcal{O}(10^{14})$, which implies the existence of inner interactions among the clones (see for instance \cite{cazenave} for experimental findings, \cite{noi1} for theoretical ones).

On the other hand, for $0.05 < \alpha+\gamma < 0.138$, by further decreasing the level of noise, one eventually crosses a line $\beta_M(\alpha+\gamma)$ below which the system develops $2P$ meta-stable retrieval states, each with a macroscopic overlap ($m \neq 0$) with some strategy. Finally, when $\alpha +\gamma < 0.05$, a further transition occurs at $\beta_C(\alpha+\gamma)$, such that below this line the single-strategy retrieval states become absolute minima of the free-energy.

\section{Poly-clonal activation: formalization and outcomes}
The system we described via the Hamiltonian (\ref{eq:H}) is actually always subjected to external
stimuli (viruses, bacteria, tumoral cells) on the effector branches. As a response to this ``work" made on the system, the mean activation of effectors may vary. Adiabatically (which is the correct limit as we are working in equilibrium statistical mechanics), this can be modeled by assuming drifted Gaussian distributions for the activity of B and CD8+ cells, that results in shifting their mean activity levels from zero to positive values $b_0$ and $k_0$, respectively. As the theory is symmetric under the switch $b_{\nu} \leftrightarrow k_{\mu}$, for the sake of simplicity (and with ALPS scenario in our mind) we can focus only on the CD8+ ensemble and we write
\begin{eqnarray}
P(\textbf{k}) \propto \exp \left( - \frac{\sum_{\mu=1}^K k_{\mu}^2}{2} \right) \Rightarrow \tilde{P}(\textbf{k}) \propto \exp \left( - \frac{\sum_{\mu=1}^K (k_{\mu}-k_0)^2}{2} \right),
\end{eqnarray}
where we used the bold style to denote a vector.
Notice that, for simplicity, we assumed that \emph{all} clones feel a stimulus, regardless of their specificity and this corresponds to a mathematical representation of poly-clonal activation.
As a consequence, the partition function (\ref{partition}) turns
out to be \be Z_{H,B,K}(\beta) = \sum_{h}\int
\prod_{\nu}^B d b_{\nu} e^{-\sum_{\nu}^B b_{\nu}^2/2} \int \prod_{\mu}^K
d k_{\mu} e^{-\sum_{\mu}^K(k_{\mu}-k_0)^2/2}e^{-\beta
\mathcal{H}_{H,B,K}(h,b,k;\xi)}. \ee By introducing the change of variables
$y_{\mu}=(k_{\mu}-k_0)$, we can solve the Gaussian integral and notice that this
maps our original system into one described by the following
Hamiltonian
\be\label{equivalente} \beta \tilde{H}(h;\xi,\Phi) =  \frac{\beta}{H} \sum_{i<j}^H \Big( \sum_{\mu=1}^K \xi_i^{\mu}\xi_j^{\mu} + \sum_{\nu=1}^B
\xi_i^{\nu}\xi_j^{\nu}\Big)h_i h_j +  \sqrt{\beta} \Phi
\sum_{i=1}^H \chi_i h_i, \ee where $\Phi \equiv \sqrt{\gamma} k_0$ is a properly rescaled measure of the mean activity, and $\chi_i =
\frac{1}{\sqrt{K}}\sum_{\mu}^K \xi_i^{\mu}$ is a random field, which in the infinite system size limit converges to a standard Gaussian
$\mathcal{N}[0,1]$.
From a statistical mechanics point of view we know that the system with shifted Gaussians for the set of variables $k_{\mu}$ can be recast into the previous one plus an external random field acting on the clones $h_i$ and whose strength is set by $k_0$, namely the width of $\mathrm{T_K}$ clonal expansion; otherwise stated, the stimulation of an effector branch acts as a perturbation on helper activities.

Actually, it is easy to see that a random field also emerges in the presence of a monoclonal activation, elicited by a given antigen, and, in general, it is able to yield a small activation of arbitrary clones, possible against self. This picture provided by our model is in agreement with the existence of low-titer self-antibodies also in hosts not affected by autoimmune diseases. Indeed, the concentration of such byproduct antibodies allows to determine a reference threshold to discriminate between low and significant activation. 

There is another deep implication in the transformation $\mathcal{H} \rightarrow \tilde{\mathcal{H}}$, induced by the stimulation: A non-negligible activation of B or CD8+ lymphocytes (i.e. $b_0$ and/or $k_0 \neq 0$) necessarily generates some sort of disorder (i.e. $\chi$) within the system.
Indeed, on the one hand we have an \emph{organized} immune response due to effector activation, on the other hand we have an \emph{disorganized} immune response due to the emergence of a random perturbation on helper branch. Interestingly, this encodes a basic thermodynamical prescription (close to the second principle) in
the framework of theoretical immunology: an \emph{ordered} work can
not be accomplished without introducing some sort of \emph{disorder} inside the system and, the larger the former the higher the level of noise introduced (note that the Hopfield network ``naturally'' represents the internal energy contribution).

It is worth remarking that, consistently with this thermodynamical picture, in physics the energy is coupled with time $t$ and, typically, ordered energy flows linearly with time ($\sim t$), while heat (disordered energy) diffuses ($\sim \sqrt{t}$). In complete analogy, by looking at eq. (\ref{equivalente}), we notice that the internal energy is coupled to $\beta$, while the heat source with $\sqrt{\beta}$.

Finally, we stress that the same formalism still holds for lymphocyte suppression, where activation is shifted towards negative values, again yielding the emergence of a random field which deranges the immune performance. Indeed, too low levels of activity of effector cells would yield to a lack of communication among them with consequent falling off of systemic regulation; otherwise stated a non-null activation level is necessary to maintain a network, that is to encode information \cite{a34,pereira,pereira2}.

\subsection{The statistical mechanics analysis of autoimmunity}

In this section we want to investigate how too strong an activity $k_0 \gg 0$ (lymphocytosis) can possibly determine pathological degenerations in the system under consideration (autoimmunity). As we are going to show, if the activation $k_0$ is
too massive, the random-field term in Eq. (\ref{equivalente})
prevails against the Hopfield interaction term (responsible for strategy retrieval) such
that the system behaves essentially randomly, inducing wrong signalling among CD4+ cells and other lymphocytes, and consequently auto-immunity.
In general, the ability of the system to retrieve stored patterns depends on the parameter set $(\beta, \alpha+\gamma, \Phi)$: The mutual balance between such quantities determines whether, in the presence of a stimulation, the system succeeds in properly cope with it according to what learned in the past.

The statistical mechanics solution of the model is rather technical and details are left to  Appendix B, while here we sketch the main results.
At first, in order to get familiar with the model, we consider a very simple situation where the antigen is detected only by, say, lymphocyte $b_{1}$, so that we simply focus on the retrieval of the first \emph{pure state}, that is we look at the regions, in the ($\alpha,\beta,\Phi$) space, where only one generic  Mattis magnetization, i.e.  $m_1=m$, may increase (for suitably initial conditions), while all the others remain zero. Beyond $m$, another parameter which turns out to be useful is $q  \equiv \mathbb{E} {(1/N) \sum_{i=1}^H  \omega(h_i)^2}$, which measures the spin-glass weight \cite{amit}.
Exploiting replica trick techniques \cite{peter,MPV}, the free energy of the system is found to be
\begin{eqnarray} \label{RSfreeenergy}
f(\alpha,\beta,\gamma,\Phi; m,q) &=& -\frac{\log 2}{\beta}+\frac{\alpha+\gamma}{2}[1+\beta r
(1-q)] + \frac12 \beta \sum_{\mu}m_{\mu}^2 + \\ \nonumber &+&
\frac{\alpha+\gamma}{2\beta}\Big[
\log[1-\beta(1-q)] -\frac{\beta q }{1-\beta(1-q)} \Big] \\
&-& \frac{1}{\beta}\langle \int d\mu(\eta) \int d\mu(z) \log 2
\cosh \Big( \beta m +  \frac{\beta \sqrt{(\alpha + \gamma) q}}{1 - \beta(1 - q)} z
 + \sqrt{\beta}\Phi\eta \Big)\rangle_{\xi}, \nonumber
\end{eqnarray}
where we fixed $m_{\mu}=m(1,0,0,...,0)$.
\newline
Extremizing again the replica symmetric free energy $f(\alpha,\beta,\gamma,\Phi; m,q)$ with respect to $m,q$, we can find
the self-consistent relations
\begin{eqnarray} \label{RSmq} 
m &=& M(\alpha,\beta,\gamma,\Phi; m,q) = \langle \xi \int d\mu(\eta) \int d\mu(z) \tanh \xi^{\mu}\Big(\beta m + \frac{\beta \sqrt{(\alpha +\gamma) q}}{1 - \beta(1 - q)} z
 + \sqrt{\beta}\Phi\eta  \Big)\rangle_{\xi}, \\ \label{RSqm}
q &=& Q(\alpha,\beta,\gamma,\Phi; m,q) = \langle \int d\mu(\eta) \int d\mu(z) \tanh^2\Big(\beta m + \frac{\beta \sqrt{(\alpha + \gamma)q}}{1 - \beta(1 - q)} z
 + \sqrt{\beta}\Phi\eta \Big)\rangle_{\xi}
\end{eqnarray}
Now, for a given set of parameters $\alpha,\beta,\gamma,\Phi$, the values of such observables allow to understand whether the retrieval can be successful, thus we solve numerically (details can be found in appendix C) Eqs. (13-14). In general, we find that $\forall \beta > 1$ and $\Phi \geq 0$, there always exists a solution with $m=0$ and $q>0$, which corresponds to a spin-glass phase.
Beyond such a solution, a pure state solution ($m>0$) appears below a critical noise $\beta_M(\alpha+\gamma,\Phi)$ and in order to discriminate which is the more stable solution (between the pure state and the spin glass), we compared the relative free-energies to look for the lowest, finding that at relatively large noise the pure state is not stable, that is it is only a local minimum; by further decreasing the noise $\beta$, the pure state becomes a global minimum.

Hence, similarly to what happens in the traditional Hopfield model ($k_0=0$), the amplitude of the pure state appears in a discontinuous way as far as the noise is lowered below a certain point, which defines a critical line $\beta_M(\alpha+\gamma, \Phi)$, but only when the noise is further lowered below a certain point, that defines a second critical line $\beta_C(\alpha+\gamma, \Phi)$ the pure state become the lowest free energy state that is, a global minimum.
Results are summarized in the phase diagrams of Fig.~$2$.

We recall that, in our framework, the pure retrieval phase represents the exposition of immune system to a particular antigen (only one particular activation pattern is retrieved) and summarizes the simplest case. Beyond this, one can also consider \emph{spurious states}:  For instance, a "spurious state" with three strategies of activation will be described by three Mattis magnetizations $m_1 \neq 0, m_2 \neq 0, m_3 \neq 0$, while the remaining are vanishing, that is $\mathcal{O}(\sqrt{H}^{-1})$ at finite volume (and zero in the infinite system size limit).
Spurious (or mixed) states represent the ability of the system to follow multiple paths of cytokines activations at the same time, interestingly turning the large spurious land of these associative models as the main interesting part in this context. Of course, one could solve numerically the set of self-consistency equations for amplitude of mixture states under whatever ansatz, calculations are just more complex and lengthy and we plan to report soon on this investigation.

\begin{figure}[tb]\label{fig:diagrammi}
{\includegraphics[width=7cm]{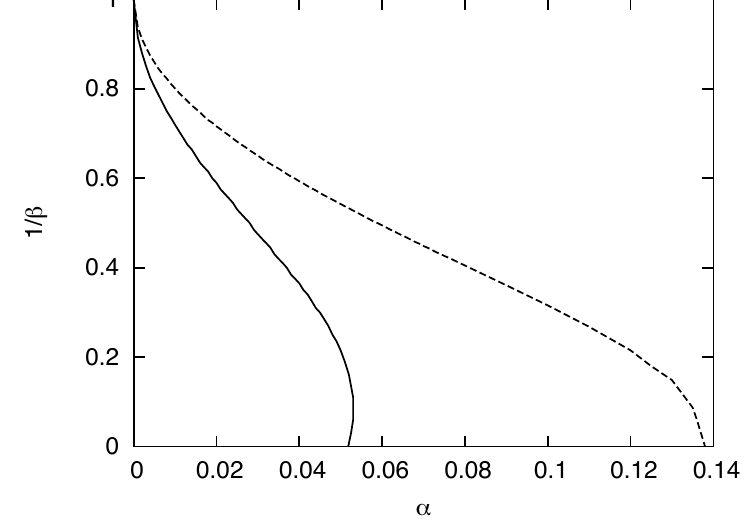}}
{\includegraphics[width=7cm]{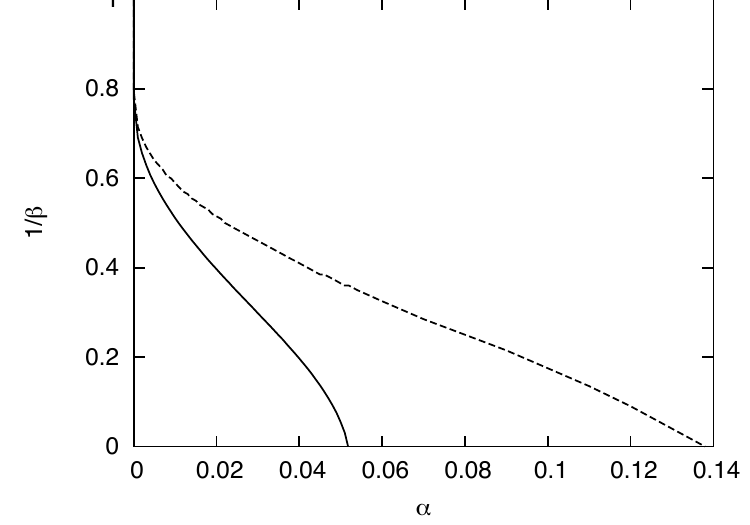}}
\\
{\includegraphics[width=7cm]{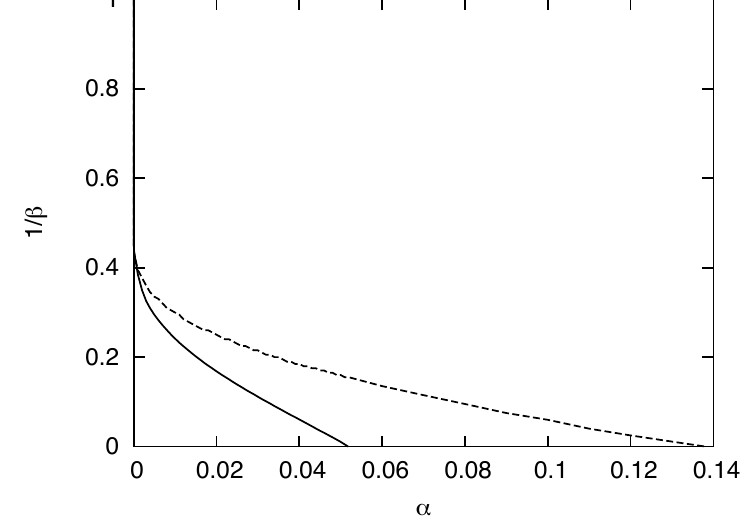}}
{\includegraphics[width=7cm]{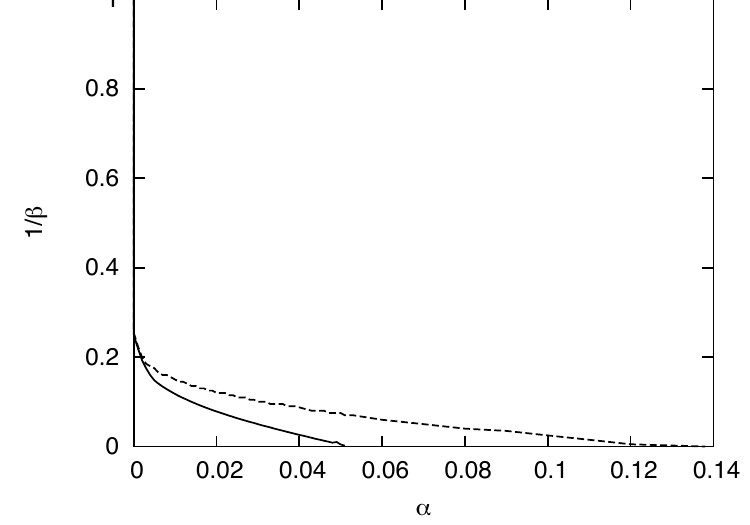}}
\caption{Phase diagrams: the dashed line represents the critical line $\beta_M$, which distinguishes among retrieval (in general sense) and spin glass phases, while the continuous line represents the critical line $\beta_C$, which confines the pure state phase. Upper panel: $\Phi=0$ (left) and $\Phi=0.5$ (right); Lower panel: $\Phi=1$ (left) and $\Phi=1.5$ (right).}
\label{fig:Diagramma_di_fase_color_linea_TC_TM_phi_0.0}
\end{figure}

\subsection{ALPS, HIV and Ageing}
The schematic representation of Fig. $3$ shows that there are basically three ways to escape from the performing region: By increasing the extent of activation $\Phi$ (toward a Random Field Phase), by breaking the balance of the ratios among different lymphocytes $B/T_{\mathrm{H}},T_{\mathrm{K}}/T_{\mathrm{H}}$ (toward a Spin Glass Phase), or by increasing the level of noise $\beta$ (toward a Paramagnetic Phase).
Interestingly, all these scenarios can be easily related to well-known conditions:
\begin{itemize}

\item Example of the random field escape: Lymphocytosis.

Autoimmune lymphoproliferative syndrome (or similar variants, see e.g. \cite{varia1,varia2}) arises in people who inherit mutations
in genes that mediate T-lymphocyte apoptosis, which is
fundamental for the immune homeostasis (a healthy steady state for the host), by
limiting lymphocyte accumulation and minimizing reactions against self-antigens. As a
result of inefficient apoptosis, lymphocytes grow monotonically in time and already in childhood severe autoimmune phenomena appear \cite{infants}.
\newline
Such phenomenology emerges consistently within our model: In the presence of a large activation of lymphocytes, the random field phase prevails against the retrieval one; intuitively, the broad range of active killer  lymphocytes makes helper cells to secrete arbitrary amounts of cytokines, ultimately loosing any capability of synergy.

\item Example of the spin-glass escape: Chronic Infection.

An imbalance between the relative sizes of the subpopulations determines an increase of the parameters $\alpha,\gamma$ and this may yield the system far from the retrieval region. Note, however, that the system now degenerates into a spin glass, a different scenario with respect to the previous case (random field). These different zones in statistical mechanics correspond, in fact, to different immunological complications: While the former corresponds to an autoimmune manifestation, the latter is close to the well known transition from simple HIV infection to the overt AIDS disease. In fact, HIV infects and kills helpers  decreasing monotonically in time the amount of these cells and consequently increasing the $\alpha,\gamma$ values.
A similar effect occurs in the presence of EBV infection since it somehow immortalizes B cells, with consequent anomalous increase of $\alpha$. We stress that, although in both cases the net effect is a spin-glass escape from the retrieval region, the causes are completely different. The whole picture gives a rationale for the understanding of the CD4+/CD8+ ratio inversion phenomenon\footnote{Notice that diluting the network towards more realistic descriptions $\alpha_c$ approaches $1$ \cite{sher} in agreement with the experimental findings \cite{abbas}.}.

\item Example of the paramagnetic escape: Ageing.

In our framework the causes of ageing can be free-radicals, by products, molecular cross-linking, damage accumulation and so on, which may preclude a firm binding between molecules and/or a slowdown in recognition processes \cite{age1,age2}. Even though we have only heuristic arguments, this kind of aging can be bridged with the ``real'' aging of the living host. It is also interesting to notice that a smaller value of $\beta$ (larger disorder) makes the critical level for $\Phi$ smaller, consistently with the well-known correlations found in ALPS patients, that is, the risk of neoplastic complications grows with patient's age \cite{infants}.

\end{itemize}

\begin{figure}[tb]\label{Uovo}
\begin{center}
{\includegraphics[width=9cm]{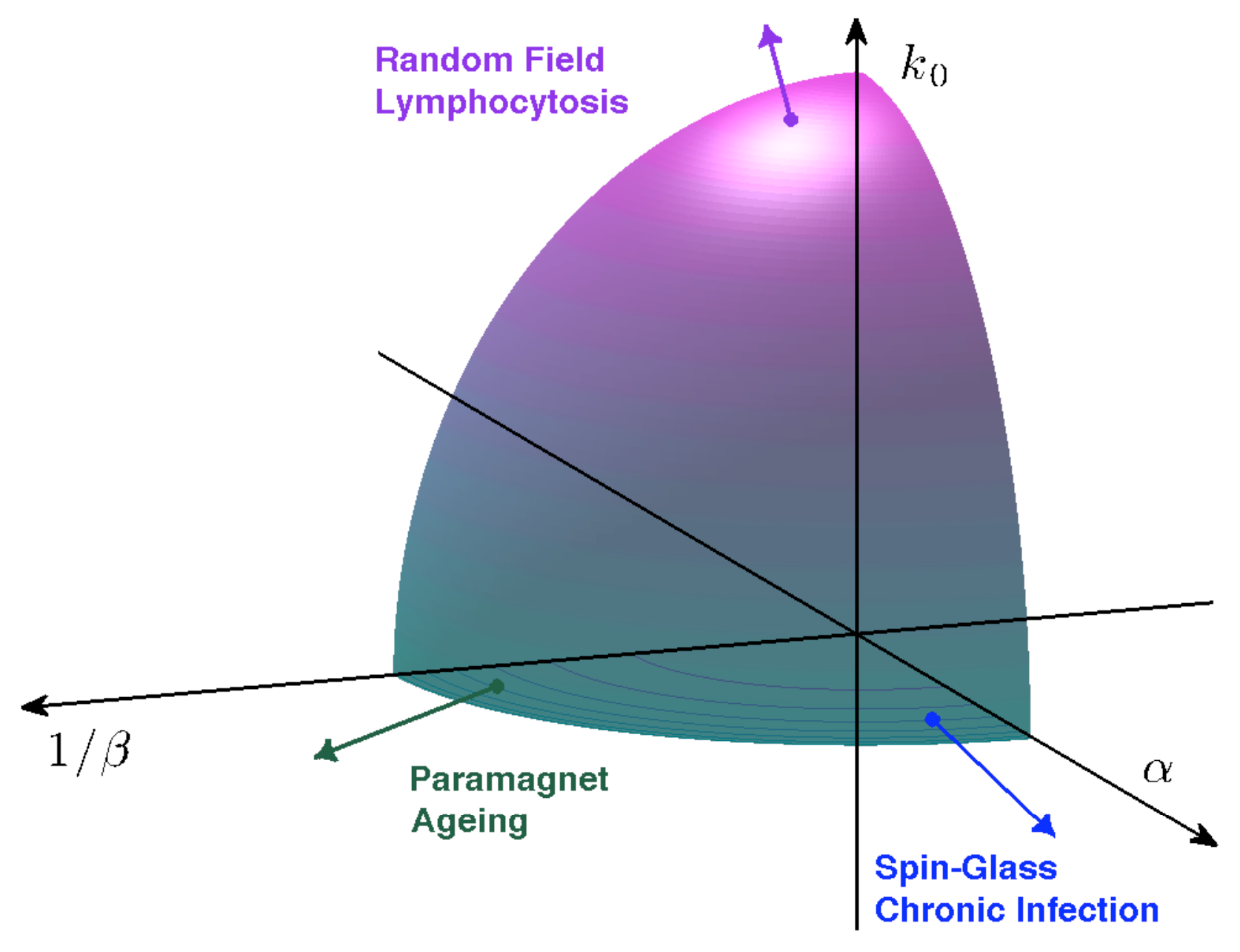}}
\caption{Schematic representation of the retrieval region. The three normal ways for escape have been depicted: Random Field (too large or too low mean activity), Spin-Glass (too large relative number of patterns) and Paramagnetic (too large degree of noise). Such states correspond to unhealthy situations, namely lymphocytosis, chronic infections and senescence, respectively. Notice that in this plot the quantity  $\alpha+\gamma$ has been renamed $\alpha$, consistently with the appendices, and that, the retrievial region is restricted, by definition, to the quadrant $\alpha>0$, $\beta>0$.}
\end{center}
\end{figure}

\section{Outlook}
In this work we introduced and analyzed a model to describe the mutual interactions occurring among lymphocytes via cytokine exchanges. While the activity of helper cells is described by a dichotomic variable $h_i = \pm 1$ (where $i$ denotes the specificity of the clone), the activity, or clonal extent, of B and killer cells is described by continuous, Gaussian-distributed variables denoted as $b_{\nu}$ and $k_{\mu}$, respectively:  This choice is a result of a relaxation of an opportune Ornstein-Uhlembeck process where the interactions among agents belonging to the same branch ($B-B$, and $T_K-T_K$) create a quadratic self-interaction term in a mean-field approximation which implies Gaussian distribution in their equilibrium values.
Interestingly, this result allows to bypass a direct estimate of the number of cells which means we can avoid dealing with chemical potentials and grand-canonical environments; in fact, we can let the clonal activity vary still retaining a canonical framework by considering not the cells, but the clones as quasi-particles.
As for cytokine, the inhibitory/excitatory function of messages they carry from helper $i$ on the effector clone $\mu$ is encoded by $\xi_i^{\mu}=-1 /+1$. The whole ensemble is then formalized by a tripartite spin-glass system where CD4+ cells can interact with the so called effector branches, i.e. CD8+ cells and B lymphocytes via these exchanges.

Firstly, we showed that such a system is equivalent to a neural network where stored patterns correspond to strategies to fight against antigens and are possibly memorized during previous infections. More precisely, cytokines patterns work as synapses providing Hebbian-like interactions
among helpers: Helpers effectively behave as an associative neural network able to store and retrieve
specific strategies in cytokine secretions for effector regulations.

Then, we have mimicked the occurrence of a (poly-clonal) activation of the branches by shifting the Gaussian distribution of these effector lymphocytes to a non null value: We proved that a state of clonal expansion (lymphocytosis) or, similarly, of clonal suppression (immunodeficiency), is formally equivalent to a random field acting on helper cells; obviously, if the strength of such a field prevails against the normal interactions, the immune system would not work correctly, possibly giving rise to autoimmune diseases. 
Such a mapping also reveals a kind of immunological version of the second principle of thermodynamics: An ordered work (clonal expansion/suppression) can
not be accomplished without introducing some sort of disorder, here random fields, inside the system itself (possibly giving rise to low-titer self-antibodies).

We also performed an analytical study of the model via the replica trick obtaining (at the replica symmetric level) a set of self-consistent equations for the order parameters as  functions of the parameters $\beta$ (degree of noise), $\alpha$ (repertoire width ratio) and $\Phi$ (clonal expansion extent). The numerical solution of these equations allows to build up a phase diagram for the performance of the system; in particular, we found that there is a region in the space $(\beta, \alpha, \Phi)$, where
helper lymphocytes can correctly work as an associative network. The system may escape from this healthy state in
three ways: Unbalancing
the amount of the relative sizes of the lymphocytes (e.g. HIV and EBV infections where the
formula CD4+/CD8+ is reversed), performing too strong a response, namely a lymphocytosis (e.g. ALPS) or simply increasing the level of white noise (e.g. ageing of the system).
The correlation between lymphocytosis and autoimmunity, or between ageing and autoimmunity, which still lacks a complete justification, at least in terms of statistical mechanics, finds here a clear explanation.

Furthermore, from these results we can also observe that a performing immune system is a system able to spread information over the network of cells: In fact, as $\alpha$ can not exceed a threshold, the system must be able to respond to a large number of different antigens with the smallest possible repertoire, which is in agreement with every systemic observation on the immune network \cite{b1}. A performing system must disentangle in multiple pattern recognitions via spurious states, minimizing in this way  the amount of required antigenic information for binding, in agreement with  the experimental findings \cite{antonio,cazenave,tony}.
Along this line we notice that such network can be established only in the presence of some intrinsic activity: If $k_0$ or $b_0$ is too low, no connection among lymphocytes can be established and no information can be spread \cite{a34,pereira,pereira2}.

A considerable amount of research is clearly opened: First of all, one could investigate about the lack of symmetry in the interaction matrix because, so far, we assumed
that the signal sent by a given helper to, say, a given killer is the same as the other way round. Despite in physics (especially in equilibrium statistical mechanics) this symmetry (the third law of dynamics)
ensures convergence to the Maxwell-Boltzmann equilibrium, in biology this property does
not hold straightforwardly. However, once related the capacity of the network with the ratios among
different types of lymphocytes, and properly introduced the clonal expansion, the whole
approach \`{a} la Gardner \cite{gardner} for unbalanced neural networks applies and could bring us closer
to the quantitative world.

\section*{Acknowledgements}
\noindent
This work is supported by FIRB grant RBFR08EKEV.
\newline
The authors are pleased to thank Alberto Bernacchia, Raffaella Burioni, Bernard Derrid\`{a}, Silvio Franz, Giancarlo Ruocco, Anna Tramontano and Guido Valesini for useful discussions.
The authors are also indebted to Cristina Cerboni for precious discussions on immunological issues.

\section*{Appendix A: The hidden role of interactions in the Gaussian activities}

In this appendix we provide an explanation about the Gaussian activity assumption we made for the effector  branches $B, T_K$; our arguments are based on the idea that their activity is also regulated by intrinsic interactions within each branch, which can be looked at in terms of an interaction network connecting B cells and $\mathrm{T_K}$ cells, respectively. Indeed, it has been evidenced that both cells exhibit, even in the absence of antigenic stimulation, a non-null activity which allows the maintenance of a network of mutual regulation even in rest conditions \cite{a34,pereira,pereira2}. From a mathematical point of view, an analogous behavior for B and $\mathrm{T_K}$ cells is intrinsic in our model due to the symmetry under the change $B \leftrightarrow T_K$.

We summarize the underlying mechanisms focusing only on B cells. Now, B lymphocytes secrete antibodies, which, given the huge amount of different clones, may detect antibodies secreted by other lymphocytes: Via this mechanism,
antibodies not only detect antigens, but also function as
individual internal images of certain antigens and are themselves
being detected and acted upon. In this way an interaction network for B cells is formed and it provides
the immune system with a "dynamical memory", by keeping the
concentrations of antibodies and of lymphocytes at appropriate levels.
\newline
This part of theoretical immunology (early experimentally investigated in \cite{antonio,cazenave,tony}, then formalized in \cite{parisi,perelson,perelson2} and within a statistical mechanics framework in \cite{noi1,noi2} can  be intuitively understood as follows: At a given time an antigen is introduced in
the body and starts replication; let us consider, for simplicity, a virus as a string of
information (i.e. $1001001$). At high enough
concentration, the antigen is detected by the proper B-lymphocyte counterpart (producing
the antibody Ig1, which can be thought of as the string
$0110110$), which then starts a clonal expansion and will release high levels of
Ig1. As a consequence, after a while, other B-cells with a consistent anticorpal affinity with Ig1 (say $1001011$, $1001000$) will meet it and, as this string never (macroscopically) existed
before, attack it by releasing the complementary string  $1001000$ and $1001011$, that,
actually, are spurious copies (internal images) of the original virus but
with no DNA or RNA charge inside: The interplay among such antibody concentrations keeps
memory of the past infection and allows a network of mutually interacting lymphocytes whose topological properties have also been shown to be able to explain basic phenomena such as self/non-self recognition and low-dose tolerance \cite{noi1,noi2}. Another issue following this point is that, while B cells of the same clone do not interact among themselves, B cells belonging to different clones (provided that their anticorpal  matching is strong enough) tend to imitate reciprocally: if the first clone undergoes clonal expansion, the clone corresponding to the anti-antibody will follow it, and viceversa.

Let us now build a dynamical system where B cells and killers interact with each others in this way (i.e. ferromagnetically) and with helpers via cytokine exchange as well: the evolution of their activity then follows an Ornstein-Uhlenbeck process \cite{ornst} like
\begin{eqnarray} \label{eq:diff1}
\tau \frac{d b_{\nu}}{dt} = -\frac1B \sum_{l=1}^B J_{\nu l}^{(b)} b_{\nu} - \frac{\beta}{\sqrt{H}}\sum_{i=1}^H \xi_i^{\nu} h_i - \sqrt{\tau} \eta_{\nu}^{(b)},\\
 \label{eq:diff2}
\tau \frac{d k_{\mu}}{dt} = -\frac1K \sum_{l=1}^K J_{\mu l}^{(k)} k_{\mu} - \frac{\beta}{\sqrt{H}}\sum_{i=1}^H \xi_i^{\mu} h_i - \sqrt{\tau} \eta_{\nu}^{(k)},
\end{eqnarray}
where $\tau$ represents the typical time-scale for the B-cells and killers diffusion (for simplicity is taken the same for both kinds of lymphocytes), $J$ is the coupling among cells themselves, $h_i$ is the activity of the $i$-th helper, $\xi_i^{\mu}$ is the cytokine message between the effector agent $\mu$ and the helper $i$, while $\eta$ accounts for a standard white noise. Equations \ref{eq:diff1} and \ref{eq:diff2} state that the rate of change for the $\nu$-th clone belonging to B ($\mathrm{T_K}$) cells population is proportional to the stimulation provided by other B ($\mathrm{T_K}$) cells via antibody exchange (via direct contact through T cell receptors) and by $T_{\mathrm{H}}$ cells via cytokines exchange, in the presence of white noise.
\newline
Given the ferromagnetic (imitative) nature of the interactions among analogous cells (if present), i.e. $J_{\mu \nu}\geq 0 \ \forall \mu,\nu$, we can properly rescale the parameters by the mean interaction $\langle J \rangle$ and, under a mean-field assumption $\sum_{\mu}J_{\nu \mu}^{(b)} b_{\nu}/ (\langle J \rangle \, B)= b_{\nu}$, we can rewrite the process as
\begin{eqnarray}\label{process}
\tau' \frac{d b_{\nu}}{dt} = -b_{\nu} - \beta' \sum_{i=1}^H \xi_i^{\nu} h_i - \sqrt{\tau''} \eta_{\nu}^{(b)},\\
\tau \frac{d k_{\mu}}{dt} = -k_{\mu} - \beta' \sum_{i=1}^H \xi_i^{\mu} h_i - \sqrt{\tau''} \eta_{\nu}^{(k)},
\end{eqnarray}
where $\tau' \langle J \rangle =\tau$, $\beta'  \langle J \rangle \sqrt{H} = \beta$ and $\tau'' \langle J \rangle^2 = \tau$.
The right-hand-sides of Eqs.~$17-18$ can be looked at as the forces eliciting the dynamic process, hence the related potentials read as (using the bold symbol to mean a vector)
\begin{eqnarray}
V(\bold{b})=-\sum_{\mu} b_{\nu}^2/2 - \beta \sum_{i,\nu} \xi_i^{\nu}h_i b_{\nu},\\
V(\bold{k})=-\sum_{\mu} k_{\mu}^2/2 - \beta \sum_{i,\mu} \xi_i^{\mu}h_i k_{\mu}.
\end{eqnarray}
In this way, if we assume a Gaussian distribution $\mathcal{N}[0,1]$ for the activity of B and $\mathrm{T_K}$ cells, the overall system can be described by means of the Hamiltonian $H=-(1/\sqrt{H})\sum_{i,\nu} \xi_i^{\nu}h_i b_{\nu} - (1/\sqrt{H})\sum_{i,\mu} \xi_i^{\mu}h_i k_{\mu}$, exactly the one introduced in our approach (see Eq.~$1$).

Therefore, the network approach developed in the last decades and based on the mutual interaction among B lymphocytes and the tripartite system approach introduced here turn out to be in perfect agreement: in particular, within our merging the two possible mechanisms for the implementation of memory (long-lived cells or clone-clone activation in a network \a la Jerne) can be regarded no longer as competing but rather as synergic. 

We stress that, at this stage, the detailed form of the antibody matrix $J_{\mu \nu}$ does not matter (see e.g. \cite{noi1} for details), the key ingredient being only its positive definiteness (given by the imitative nature of B and $\mathrm{T_K}$ cells), to ensure its mean value $\langle J \rangle$ to exist strictly positive.
If interactions among B lymphocytes and among T lymphocytes were both inhibitory and excitatory, i.e. $J$ is non-positive-definte like in spin-glass systems, then convergence would not hold in general and we would not be able to merge the two approaches.

\section*{Appendix B: The replica trick calculation for the evaluation of the free energy}

The system, whose thermodynamics we want to tackle, is ruled by the following {\em Hamiltonian}\footnote{Strictly speaking, in standard statistical mechanics, the noise level plays a uniform role on the interactions, while we face with its linear coupling to the Hopfield terms and a square root one to the random field term. The need of a uniform influence is of course nor a biological must neither a mathematical restriction and can then be easily relaxed.} (see Eq.~\ref{equivalente}):
\be \label{H_appe} \beta\mathcal{H}(h;\xi)= - \frac{\beta}{H} \sum_{i<j}^H \Big(
\sum_{\mu}^K \xi_i^{\mu}\xi_j^{\mu} + \sum_{\nu}^B
\xi_i^{\nu}\xi_j^{\nu}\Big)h_i h_j - \sqrt{\beta \gamma} k_0
\sum_i^H \chi_i h_i, \ee where $\chi_i =
\frac{1}{\sqrt{K}}\sum_{\mu}^K\xi_i^{\mu}$, in the infinite system size limit, converges to a standard Gaussian
$\mathcal{N}[0,1]$ via a standard CLT argument\footnote{Two correlated observations are needed here. First the approximation to a Gaussian may appear dangerous because for the same CLT argument the Hebbian kernel converges to a $\mathcal{N}[0,1]$, too; this actually onsets the transition from an associative behavior to a spin glass phase. However, while the Hebbian kernel in this procedure loses its peculiar organization of the phase space able to store and retrieve information, $\eta$ is a random object even without the CTL limit and its convergence to a standard Gaussian only simplifies  calculations.}.

In order to study the retrieval phase of this system trough disordered statistical
mechanics, we apply the so-called replica trick technique \cite{mezard} (under the assumption of replica symmetry) following the derivation of Coolen, Kuhn and Sollich
\cite{peter}: The idea is to force the retrieval toward a particular ensemble of patterns $l<B+K$ mimicking a reasonable dynamics toward one of the attractors which, indeed, can be identified by these $l$ patterns; then, one needs to check where, in the region of the space $[(\alpha+\gamma),\beta,\Phi]$, suitably order parameters (the Mattis  magnetizations that we introduce later) are stable and, further, where these minima are even absolute minima of the free energy, such that we have "thermodynamical stability"\footnote{We will find even a large region where only spin glass states exist and  a large mixed region where these minima exist but are local minima, the spin glass being still the global one; the latter is the "spurious states" scenario, where, despite these are not thermodynamically stable, are still of primary interest in the dynamics as are however well defined attractors with long meta-stable lifetimes \cite{amit}.}.  Of course all the "not-recalled" patterns $P-l$ now act as a quenched noise on the retrieval of the selected $l$ patterns and we know how to deal with these remaining $\xi$-terms.
\newline
Before proceeding we notice that there is permutational invariance: thinking at
the first $l$ as the retrieved patterns is completely fictitious as any set of $l$ patterns can work finely as well. As a consequence, we introduce a new symbol $P=B+K$ because, as far as no differentiation among the subclasses $H_1,H_2$ is made, the two effector branches are indistinguishable and only their total amount versus the amount of the available helpers matters.

We now properly elaborate the Hamiltonian (\ref{H_appe}) by adding a finite number $l$ of Lagrange multipliers $\lambda_{\mu}$ to the Hamiltonian so to easily express Mattis order parameters $m_{\mu} \equiv \sum_{i=1}^H \xi_i^{\mu} h_i/H$ as derivatives of the free energy w.r.t. them; more precisely, one has
$$
\mathcal{H}(h;\xi) \Rightarrow \mathcal{H}(h;\xi) + \beta \sum_{\mu=1}^l
\lambda_{\mu}\sum_i^H \xi_i^{\mu}h_i,
$$
from which, recalling $A(\alpha,\beta,\gamma)=-\beta F(\alpha,\beta,\gamma)$, one gets
$$
\langle m_{\mu} \rangle = \frac{\partial}{\partial
\lambda_{\mu}} \frac{F}{H}|_{\lambda = 0}
$$
Basically, the role of multipliers is to force to end up in the selected attractors. One has three different kinds of noise which are not part of attractors:  $\beta$, the RF and the excluded patterns $B+K-l$.
Hence, the complete Hamiltonian which we  study is
 \be \label{H_lag}
 -\beta \mathcal{H}(h;\xi,\lambda) =
\frac{\beta}{H}\sum_{i<j}^H(\sum_{\mu}^K\xi_i^{\mu}\xi_j^{\mu}+\sum_{\nu}^B
\xi_i^{\nu}\xi_j^{\nu})h_i h_j + \sqrt{\beta \gamma} \kappa_0
\sum_i^H \eta_i h_i
- \beta \sum_{\mu=1}^l \lambda_{\mu}\sum_i^H h_i
\xi_i^{\mu}.\ee
We want to solve the thermodynamics (i.e. obtain an explicit expression for the free energy and a picture of the phase diagram by its extremization) via the replica trick, which consists in evaluating the logarithm of the
partition function trough its power expansion, namely
\begin{eqnarray} \log Z
&=& \lim_{n \to 0}\frac{Z^n-1}{n} \Rightarrow \langle \log Z \rangle =
\lim_{n \to 0}\frac{\langle Z^n \rangle - 1}{n}= \lim_{n \to
0}\frac1n \log\langle Z^n \rangle.
\end{eqnarray}
This implies that in order to obtain the mean of $\log Z$ one can average $Z^n$, which is itself a partition function of $n$ identical systems which, for any given set of random variables, do not interact: these are the ``replicas''. The intensive, i.e. divided by $H$, free energy reads off as
\begin{eqnarray} \label{free_replica}
\frac{\langle F \rangle}{H} &=& \lim_{n \to 0}\frac{1}{\sqrt{\beta} H n}\log\sum_{\{h^1,...,h^n\}}\exp\langle -\beta
\sum_{a=1}^n \mathcal{H}(h^{a}; \xi)\rangle, \end{eqnarray} where we
introduced the symbol $a \in (1,...,n)$ to label the $n$
different replicas of the system (with the same quenched
distribution of the $\xi$).
\newline
By plugging Eq.~\ref{H_lag} into Eq.~\ref{free_replica} we get
\begin{eqnarray}\nonumber  \frac{\langle F \rangle}{H} &=&
\frac{\alpha+\gamma}{2}-\frac{\log2}{\beta}-\lim_{n \to 0}\frac{1}{\beta
n H}\langle \exp\Big(-\beta \sum_{\mu}^l
\sum_{a}^n[\lambda_{\mu}\sum_i^H
h_i^{a}\xi_i^{\mu}-\frac{1}{2H}(\sum_i^H
h_i^{a}\xi_i^{\mu})^2]\Big) \\ &\cdot&  \exp\Big( \sqrt{\beta \gamma}
\kappa_0 \sum_i^H\sum_{a}^n \eta_i
h_i^{a}\Big)
\rangle_{\xi}\langle \exp\Big(
\frac{\beta}{2H}\sum_{a}^n\sum_{\mu
>l}^P(\sum_i^H \xi_i^{\mu}h_i^{a})^2\Big)
\rangle_{\xi}
\rangle_{h^{a}},
\end{eqnarray} where we can linearize the quadratic exponential terms for the $\mu < l$
with the Gaussian integral, and apply for convenience the shift
$m_{a \mu}\to \sqrt{\beta H}m_{a \mu}$, as
$$
\exp \left(\frac{\beta}{2H}(\sum_i^H h_i^{a}\xi_i^{\mu})^2\right)=
\int_{-\infty}^{+\infty}d\mu(z_{a \mu})\exp \left(\frac{\sqrt{\beta}}{\sqrt{H}}\sum_i^H
h_i^{a}\xi_i^{\mu}z_{a \mu}\right).
$$
The intensive free energy can be written now as
\begin{eqnarray}
\frac{\langle F \rangle}{H} &=& \frac{\alpha+\gamma}{2} - \frac{\log
2}{\beta} - \\ \nonumber &-& \lim_{n \to 0} \frac{1}{\beta H
n}(\frac{\beta H}{2\pi})^{(\frac{nl}{2})}\int \prod_{a
\mu}^{n l} d m_{a \mu} e^{-\frac{\beta
H}{2}\sum_{a \mu}m_{a \mu}^2} \langle \exp \Big(\beta
\sum_{\mu}^l \sum_{a}^n \sum_i^H h_i^{a}
\xi_i^{\mu}[m_{a}^{\mu}- \lambda_{\mu}]) \cdot \\
\nonumber &\cdot& \langle \exp \Big( \frac{\beta}{2H}
\sum_{a}^n \sum_{\mu>l}^P (\sum_i^H
\xi_i^{\mu}h_i^{a})^2 + \sqrt{\beta \gamma} \kappa_0 \sum_{a}^n \sum_i^H
\eta_i
h_i^{a} \Big)\rangle_{\xi,\eta}\rangle_{h^{a}}.
\end{eqnarray}
We note that, as $B$ and $K$ act together as identical interacting terms,  among the $l$ retrieved patterns we do not distinguish between those from the $B$ components and the $K$ ones; it is then useful to "diagonalize the perspective" by introducing the variable $\tilde{P}$ as follows:
\newline
We have a global amount of $B+K=P$ clones. In this set, $l$ are retrieved, $P-l$ are the remaining terms. Among these, $P-l-\tilde{P}$ can be though of as the responsible for the (not retrieved) interactions with the $B$ (in the corresponding $3$-parties spin glass) and $\tilde{P}$ are left for the interactions with the $K$ (in a nutshell it is a reshuffling). We can then average over the quenched noise and 
linearize even the term $\langle \exp\Big(
\frac{\beta}{2H}\sum_{a}^n\sum_{\mu>l}^P (\sum_i^H
\xi_i^{\mu}h_i^{a})^2 \Big) \rangle_{\xi}$ so to write
the free energy as
\begin{eqnarray}
&& \frac{\langle F \rangle}{H} = \frac{\alpha+\gamma}{2}-\frac{\log
2}{\beta}- \lim_{n \to 0}\frac{1}{\beta H n} \left(\frac{\beta
H}{2\pi}\right)^{\frac{nl}{2}}\int \prod_{a, \mu}^{n, l} \cdot 1 \cdot
e^{-\frac{\beta H}{2}
\sum_{a, \mu}^{n, l}m_{a \mu}^2}
\\ \nonumber
&& \langle \exp\Big(\sqrt{\beta}\sum_{\mu<l}\sum_{a,i} h_i
\xi_i^{\mu}(m_{a}^{\mu}-\lambda_{\mu}) \Big)  \exp \Big(H (\alpha+\gamma) \log
\left[\int \prod_{a}^n d
\mu(z_{a})\exp(\frac{\beta}{2}\sum_{a,\beta}z^{a}q_{a
\beta}z^{\beta})\right]+\sqrt{\beta \gamma}
\kappa_0 \sum_{i,a}^{H,n} \eta_i
h_i^{a} \Big) \rangle_{h_i^{a}} ,
\end{eqnarray}
where the term $1$ has been introduced symbolically into the
expression so to be rewritten as \be 1=\int \prod_{ \alpha \beta} dq
\delta[q_{\alpha \beta}-\frac1H \sum_i
h_i^{a}h_i^{\beta}]= (\frac{\beta H}{2\pi})^{n^2}
\int \prod_{ \alpha \beta} dq_{\alpha \beta} \int
\prod_{ \alpha \beta} d\tilde{q}_{ \alpha \beta} \exp(i H \sum_{\alpha
\beta}\tilde{q}_{\alpha \beta}[q_{\alpha \beta}-\frac1H \sum_i^H
h_i^{\alpha}h_i^{\beta}]). \ee Now we assume the
commutation of the limits $\lim_{n \to 0}, \lim_{H \to \infty}$ and get
\begin{eqnarray} \nonumber
\lim_{H \to \infty} \frac{\langle F \rangle}{H} &=& \frac{\alpha+\gamma}{2}-\frac{\log
2}{\beta} \\ \nonumber &-& \lim_{n\to 0}\lim_{H \to \infty}
\frac{1}{\beta H n}\mathbb{E}\int \prod_{a \mu} dm_{a
\mu}\int
\prod_{\alpha \beta} d q_{\alpha \beta}\int \prod_{\alpha \beta}d\tilde{q}_{\alpha \beta} \\
&\cdot& \exp{\Big[H\Big( i \sum_{\alpha
\beta}\tilde{q}_{\alpha \beta} q_{\alpha \beta}-\frac12 \beta
\sum_{a \mu}m_{a \mu}^2 + (\alpha+\gamma) \log \int
\prod_{a}^n
d\mu(z_{a})e^{\frac{\beta}{2}\sum_{\alpha \beta}z_{a}q_{\alpha \beta}z_{\beta}}
\Big) \Big]} \\ \nonumber &\cdot& \langle \exp{\Big[ \beta
\sum_{\mu<l}\sum_{a}^n \sum_i^H
h_i^{a}\xi_i^{\mu}[m_{a}^{\mu}-\lambda_{\mu}]
-i\sum_{\alpha \beta}\tilde{q}_{\alpha \beta}\sum_i^H h_i^{a}h_i^{\beta}
+ \sqrt{\beta \gamma} \kappa_0 \sum_i^H \sum_{a}^n
\eta_i h_i^{a}\Big]}\rangle_{h^{a}}. \nonumber
\end{eqnarray}
The $n$-dimensional Gaussian integral over $z$ factorizes in a
standard way after appropriate rotation of the integration
variable as  \be \log \int
d\mu(z_{a})\exp\Big(\frac{\beta}{2}\sum_{\alpha \beta}z_{\alpha}q_{\alpha \beta}z_{\beta}\Big)
= -\frac12 \log \det [\mathbb{I}-\beta \mathbb{Q}], \ee which
allows to rewrite the free energy as
\begin{eqnarray}
\nonumber
\lim_H \frac{\langle F \rangle}{H} &=& \frac{(\alpha+\gamma)}{2}-\frac{\log
2}{\beta} - \lim_{n\to 0}\lim_{H\to \infty} \frac{1}{\beta H
n}\mathbb{E}_{\xi,\eta} \int d m_{a \mu}\int d q_{\alpha \beta}\int
d\tilde{q}_{\alpha \beta} \\
\nonumber
&\cdot& \exp\Big([H ( i \sum_{\alpha
\beta}\tilde{q}_{\alpha \beta} q_{\alpha \beta}-\frac12 \beta
\sum_{a \mu}m_{a \mu}^2)]\Big)
\exp\Big([H(- \frac{\alpha+\gamma}{2}\log
\det [\mathbb{I}-\beta \mathbb{Q}])]\Big)
\\ &\cdot& \prod_i^H \langle \exp\Big(\beta \sum_{\mu<l}
\sum_{a}^n
h^{a}\xi_i^{\mu}(m_{\mu}^{a}-\lambda_{\mu})-i[
\sum_{\alpha \beta}h^{\alpha}q_{\alpha \beta}h^{\beta}+\sqrt{\beta \gamma}
\kappa_0 \eta \sum_a^n h^{a}]\Big)
\rangle_{h^{a}},
\end{eqnarray}
where $\mathbb{E}_{\xi,\eta}$ represents the average over the quenched variables $\xi, \eta$
As we reached a formulation where all the exponents are extensive in the volume $H$, we are allowed to apply the saddle point method such that the extremal
$\langle f(m,q,\tilde{q}) \rangle \exists : \lim_{H \to 0} \langle F
\rangle_{\xi}H^{-1}=\lim_{n\to 0}\lim_{H \to \infty} \langle f(m,q,\tilde{q})
\rangle$, being
\begin{eqnarray} \nonumber
\langle f(m,q,\tilde{q}) \rangle &=&
\frac{\alpha+\gamma}{2}-\frac{\log 2}{\beta} - \lim_{n\to 0}
\frac{1}{\beta n}\mathbb{E}_{\eta}\Big[ \langle \log \langle
\exp\Big( \beta \sum_{\mu<l}
\sum_{a}^nh_{a}\xi_{\mu}(m_{\mu}^{a}-\lambda_{\mu})\Big)
\\ &\cdot& \exp\Big( -i\sum_{\alpha \beta}h^{\alpha}
q_{\alpha\beta}h^{\beta}+\sqrt{\beta \gamma}\kappa_0\eta
\sum_{a}^n h^{a} \Big) + \\
&+& i \sum_{\alpha \beta}^n \tilde{q}_{\alpha \beta}
q_{\alpha \beta}-\frac12 \beta
\sum_{a\mu}^{nl}m_{a \mu}^2 -\frac12
(\alpha+\gamma) \log\det[\mathbb{I}-\beta \mathbb{Q}]
 \Big]\rangle_{\xi}\rangle_{h^{a}}.
\end{eqnarray}

Let us study the behavior of the replica-symmetric matrix
$$
\Lambda_{a\beta}=[1-\beta(1-q)]\delta_{a\beta}-\beta
\mathbb{Q},
$$
where the matrix $\mathbb{Q}$ has all the off diagonal entries
equal to $q$ and the diagonal ones to $1$. There exist two
eigenvectors, namely $\bold{x}=(1,1,...,1)$ with algebraic
multiplicity $1$ and eigenvalue $\lambda_1=1-\beta(1-q)-\beta q
n$, and $\bold{\hat{x}}=\sum_{a}x_{a}=0$, namely the
whole hyperspace orthogonal to the first eigenvector. Of course
the algebraic multiplicity of the latter is $n-1$ and its
eigenvalue $\lambda_{\hat{1}} = 1-\beta(1-q)$. So we can write the
determinant of the matrix $\Lambda$ as the product of all its
eigenvalues to get
\begin{eqnarray}\nonumber
\log\det \Lambda &=& \log \prod_i \lambda_i =
\log[1-\beta(1-q)-\beta q n]+(n-1)\log[1-\beta(1-q)] \\ &=& n
\Big[\log[1-\beta(1-q)] - \frac{\beta
q}{1-\beta(1-q)}\Big]+O(n^2),
\end{eqnarray}
as we are expanding around small $n$ because we are approaching the
$n\to0$ limit.
\newline
Overall we can rewrite the free energy as
\begin{eqnarray}\label{cisiamoquasi} \nonumber
f(m,q,r) &=& -\frac{\log 2}{\beta}+\frac{\alpha+\gamma}{2}\Big( 1+\beta r
(1-q)\Big) + \frac12 \sum_{\mu} m_{\mu}^2 +
\\ &+& \frac{\alpha+\gamma}{2\beta}
\Big[ \log\Big(1-\beta (1-q)\Big)  -\frac{\beta q}{1-\beta(1-q)}\Big] - \\
&-& \frac{1}{\beta n}\mathbb{E}_{\eta}\langle \log \langle \exp\Big([
\beta (\sum_{a}h^{a})(\sum_{\mu<l}m_{\mu}\xi^{\mu})
+ \frac12 (\alpha+\gamma) \beta^2 r (\sum_{a}h^{a})^2 +
\sqrt{\beta \gamma}\kappa_0
\eta(\sum_{a}h^{a})]\Big)\rangle_{h}\rangle_{\xi}.
\nonumber
\end{eqnarray}
Now, focusing on the last line of the expression above we can
linearize the quadratic term $(\sum_{a}h^{a})^2$
through a standard Gaussian integral representation,
$$
\exp\Big( -\frac12 (\alpha+\gamma) \beta r (\sum_{a}h^{a})^2
\Big) = \int d \mu (z) \exp\Big( \beta \sqrt{(\alpha+\gamma) r}z
(\sum_{a}h^{a}) \Big),
$$
and get (writing once again only the last line of expression
(\ref{cisiamoquasi}))
\begin{eqnarray} \nonumber
&& -\frac{1}{\beta n}\mathbb{E}_{\eta}\langle \log \langle d\mu(z)
\exp\Big[\Big(\sum_{a}h^{a})\Big( \beta
(\sum_{\mu<l}m_{\mu}\xi^{\mu})+ \beta \sqrt{(\alpha+\gamma) r}z
 + \sqrt{\beta \gamma}\kappa_0 \eta \Big)\Big]\rangle_{h}\rangle_{\xi} = \\
&& \frac{1}{\beta n}\mathbb{E}_{\eta} \langle \log\int d\mu(z) 2^n
\cosh^n\Big( \beta
\Big((\sum_{\mu<l}m_{\mu}\xi^{\mu})+\sqrt{(\alpha+\gamma) r}z\Big)
 + \sqrt{\beta \gamma}\kappa_0 \eta \Big) \rangle_{\xi}
\end{eqnarray}
Now, using $\cosh^n(x) \sim 1 + n \log\cosh(x)$ and writing the
whole free energy we get
\begin{eqnarray}\label{AGSfree}
f(m,q,r) &=& -\frac{\log 2}{\beta}+\frac{\alpha+\gamma}{2}[1+\beta r
(1-q)] + \frac12 \beta \sum_{\mu}m_{\mu}^2 + \\ \nonumber &+&
\frac{\alpha+\gamma}{2\beta}\Big[
\log[1-\beta(1-q)] -\frac{\beta q }{1-\beta(1-q)} \Big] \\
&-& \frac{1}{\beta}\langle \int d\mu(\eta) \int d\mu(z) \log 2
\cosh \Big[ \beta \Big((\sum_{\mu<l}m_{\mu}\xi^{\mu})+\sqrt{(\alpha+\gamma)
r}z\Big)
 + \sqrt{\beta \gamma}\kappa_0 \eta \Big]\rangle_{\xi}. \nonumber
\end{eqnarray}
Extremizing again the replica symmetric free energy we can find
the self-consistent relations
\begin{eqnarray} \label{AGS1}
m &=& \langle \int d\mu(\eta) \int d\mu(z) \tanh\Big(\beta
(\sum_{\mu<l}m_{\mu}\xi^{\mu}+\sqrt{(\alpha+\gamma) r}z)
 + \sqrt{\beta \gamma}\kappa_0 \eta  \Big)\rangle_{\xi}, \\ \label{AGS2}
 q &=& \langle \int d\mu(\eta) \int d\mu(z) \tanh^2\Big(\beta (\sum_{\mu<l}m_{\mu}\xi^{\mu}+\sqrt{(\alpha+\gamma) r}z)
 + \sqrt{\beta \gamma}\kappa_0 \eta  \Big)\rangle_{\xi},\\
 r &=& q/\Big( 1 - \beta (1-q)^2\Big). \label{AGS3}
\end{eqnarray}
These equations must be solved numerically (the difficulty in the involved mathematics mirrors the sudden jumps in the order parameters values), to which the next section is dedicated.

\section*{Appendix C: Numerical solutions of the self-consistency equations}

As shown in Eq. (\ref{equivalente}), both the ratio $\alpha+\gamma$ between killers and helpers and the strength of the killer clonal expansion $k_0$ multiply the random field  at once: We use $\Phi = \sqrt{\gamma}\kappa_0$ as a single tunable parameter and we stress once more that, analogously to $P$ and $\tilde{P}$ in the previous appendix, only the total amount of branch lymphocytes matter in the ratio with the helpers (namely $(B+K)/H$), so we shift $\alpha + \gamma \to \alpha$ for the sake of simplicity.
\newline
In analogy with the standard Hopfield model, the phases where our system may show emergent cooperative behavior among its constituents are several and here we outline our strategy to detect the two (limiting) simpler cases.
\begin{figure}\label{intersezioni}
\includegraphics[width=4cm]{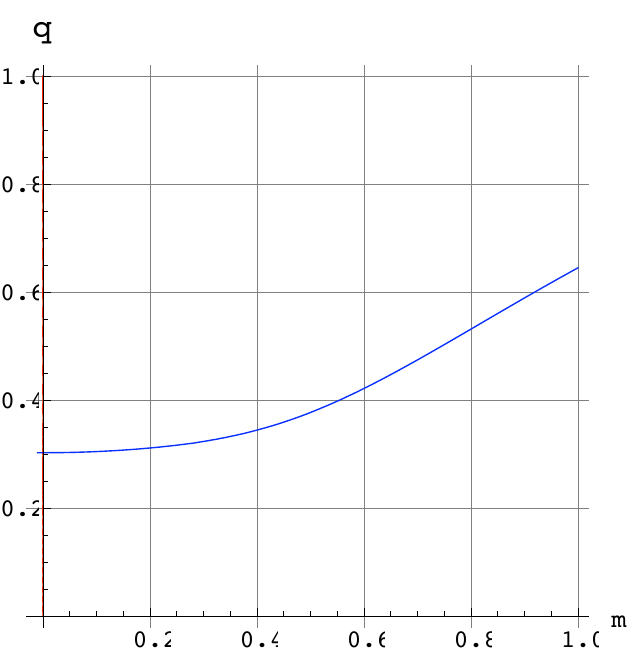}
\includegraphics[width=4cm]{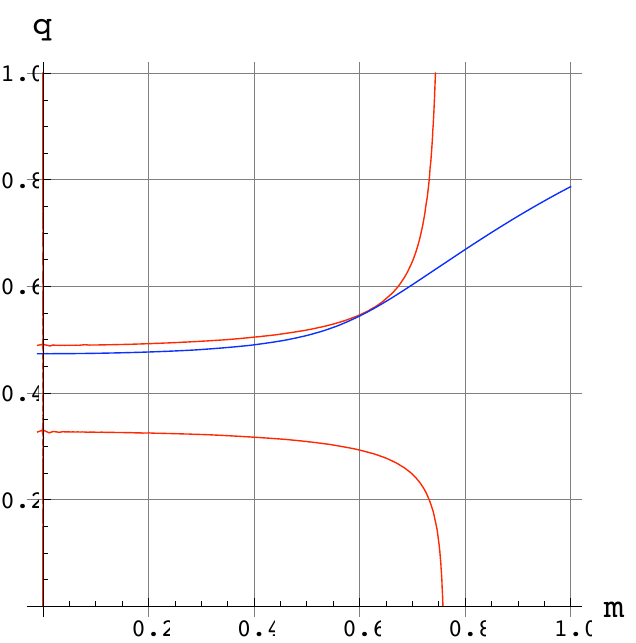}
\includegraphics[width=4cm]{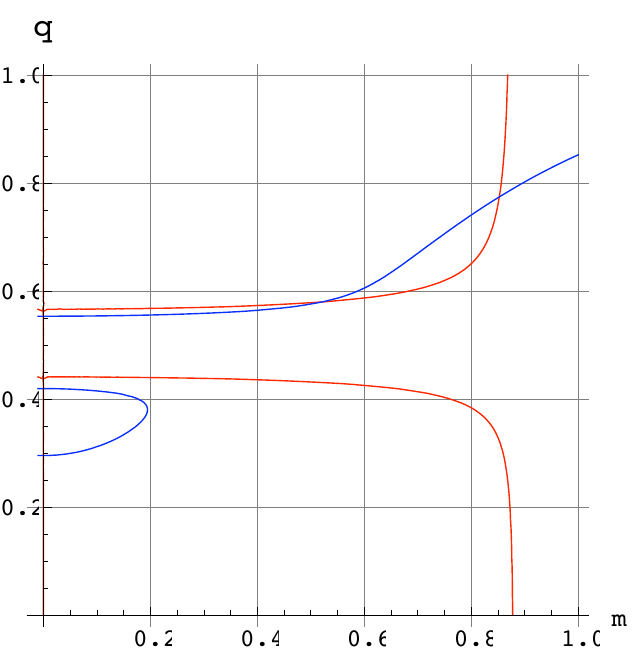}
\includegraphics[width=4cm]{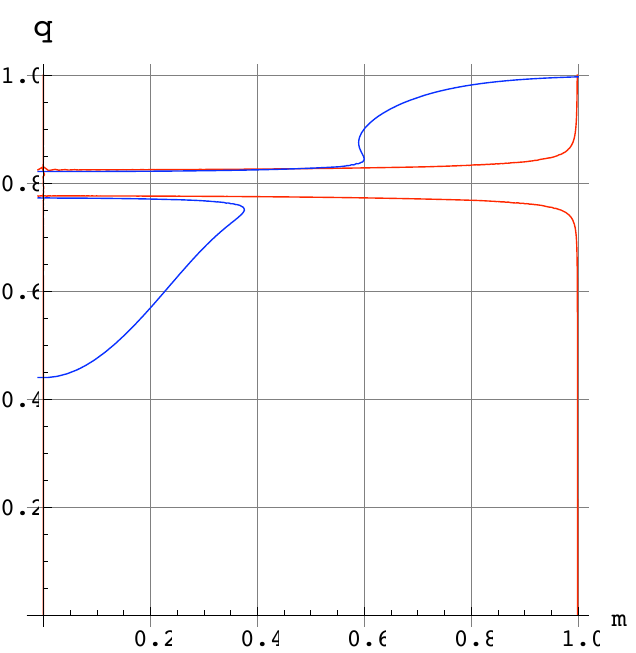}
\caption{From left to right:  Solutions of the RS self-consistency equations for $(\Phi=0.5$, $\alpha+\gamma=0.01)$. Red line: solution of $m-M(m,q,\alpha+\gamma,\Phi,\beta)=0$, Blue line: solution of $q-Q(m,q,\alpha+\gamma,\Phi,\beta)=0$.
\newline
(a) $1/\beta=0.8$. Only the upper branch counts, under the value $q=1-1/\beta$ the free energy has only complex values.
\newline
(b) $1/\beta=0.6$. In this particular point $\alpha+\gamma,\Phi,\beta^{-1}$ a pure state solution $m>0$ appears as the two contour-plot lines -for $m$ and for $q$- are tangent.
\newline
(c) $1/\beta=0.5$. Solution of the RS self-consistency equations for $(\Phi=0.5$,$\alpha+\gamma=0.01)$.
Free energy is complex along the lower branches which are therefore rejected (note that they never cross in fact). Above two intersections appear. Only the higher $m,q$ intersection is the thermodynamical pure state solution because it is coupled with the lower free energy.
\newline
(d)  $1/\beta=T=0.2$. Note that lowering the noise, (for $\alpha+\gamma<(\alpha+\gamma)_c=0.138$) we always find the pure state retrieval solution.}
\end{figure}
As for the \emph{pure states}, we look at the regions, in the ($\alpha,\beta,\Phi$) space, where only one generic  Mattis magnetization, say  $m_1=m$, may increase (for suitably initial condition), while all the others remain zero; further, with the overlap $q_{\alpha \beta}$ we can measure the spin glass weight; in fact, for high noise level ($\beta < 1$), $m =  q = 0$ and the system is ergodic (of no interest in theoretical immunology), while focusing on the low noise level ($\beta > 1$), we can distinguish a spin-glass phase with $m = 0,\ q > 0$ and a phase where the system displays associative memory with $m >0,\ q>0$.
\begin{figure}[tb]\label{bue}
\includegraphics[width=7.5cm]{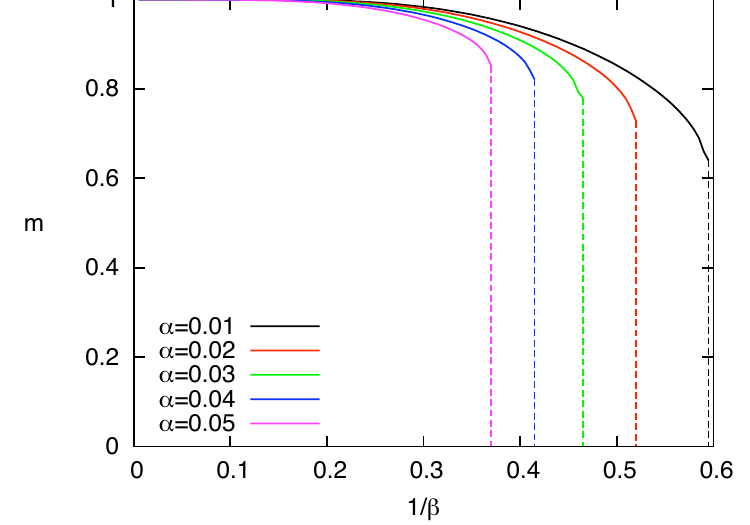}
\includegraphics[width=7.5cm]{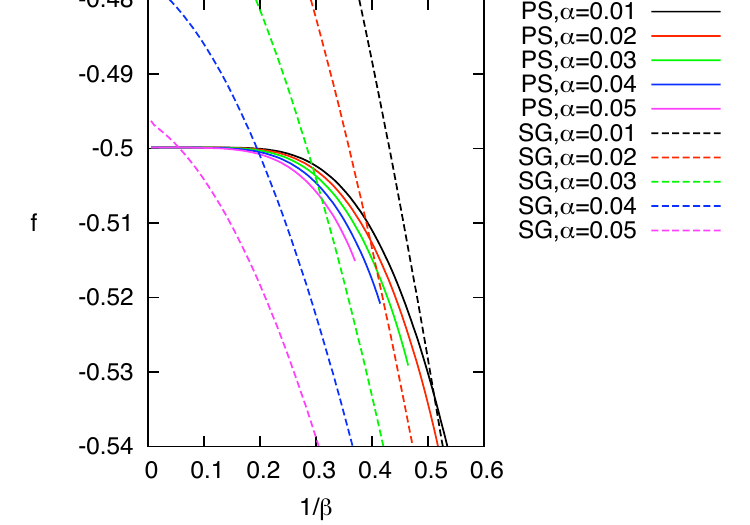}
\caption{Left: RS amplitudes of the Mattis order parameter of the pure states at $\Phi=0.5$ as function of the noise. From top to bottom: $\alpha+\gamma = 0.01 - 0.05$, $(\Delta [\alpha+\gamma]=0.01)$. Right:  Solid lines represent free energies of the pure states (PS) for $\alpha+\gamma=0.01 - 0.05$ at $\Phi=0.5$. Dashed lines represent free energies of the spin glass (SG) states for $\alpha+\gamma=0.01 - 0.05$ at $\Phi=0.5$.
Each different $\alpha+\gamma$ is called simply $\alpha$ in the plots and each couple of same lines has a different color for comparison.  The higher $1/\beta$ of the PS line defines $\beta_M$ point at each $\alpha+\gamma$. The PS and spin glass lines cross in the $\beta_C$ point for each $\alpha+\gamma$.}\label{fig:magnetizzazioni_phi_0_energie_libere_SP_SG_phi_0.5_alfa_0.01_0.05}
\end{figure}
Beyond this extremum case, there is a whole family of other cases where \emph{spurious states} appear. For instance, a "spurious state" with two patterns of activation will be described by two Mattis magnetizations $m_1 \neq 0, m_2 \neq 0$, while the remaining are vanishing, that is $\mathcal{O}(\sqrt{H}^{-1})$ at finite volume. Of course, increasing the number of antigens means increasing the $B,K$ repertoires, which lastly falls off the system toward a spin glass phase\footnote{Physically the transition to a spin-glass state is accomplished with an exponential increasing of the minima of the free energy which pushes the network into the "blackout scenario" \cite{amit}. This can be  understood intuitively as the amount of spurious states, namely linear combination of pure states (with smaller basins of attractions) grow as the Newton binomial, i.e. in a non polynomial way.}. This other extremum (the maximum amount of parallel paths of activations before collapsing into the spin-glass region) is the second case we analyze.
\begin{figure}[tb]
{\includegraphics[width=7cm]{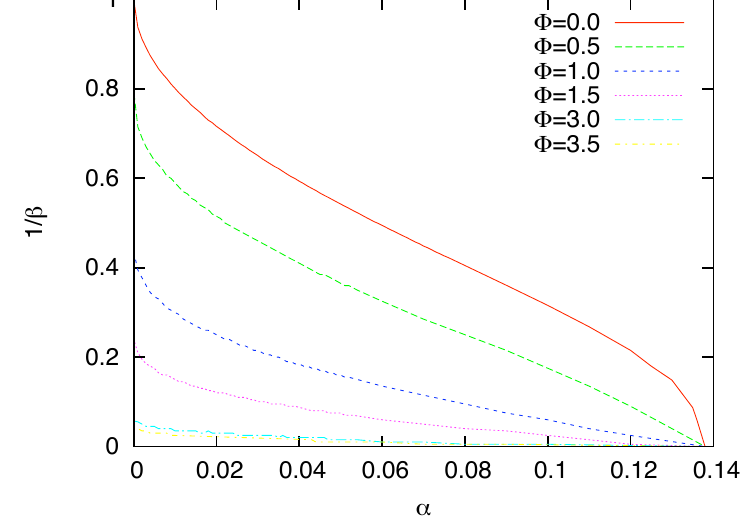}}
\includegraphics[width=7cm]{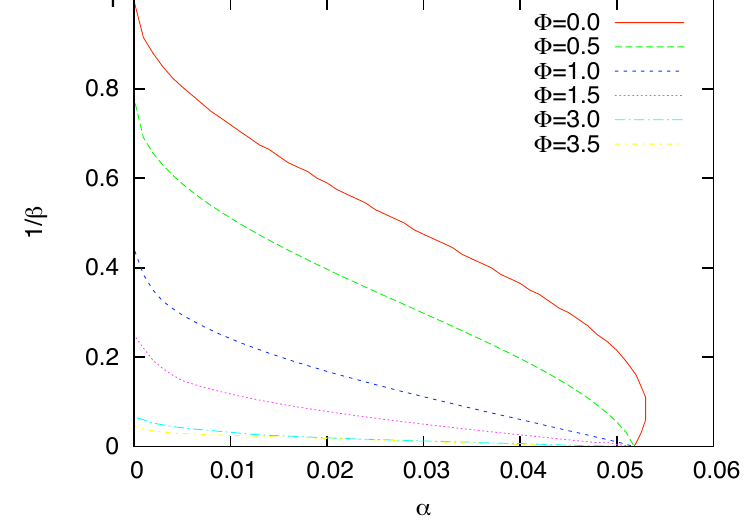}
\\
\includegraphics[width=7cm]{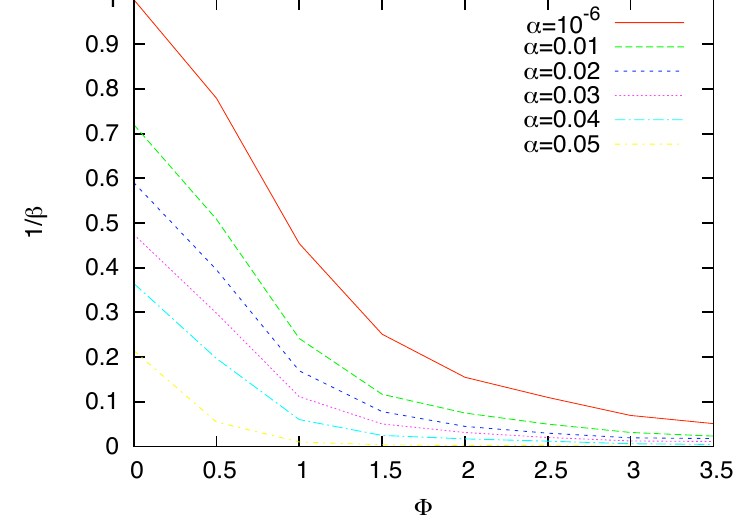}
\includegraphics[width=7cm]{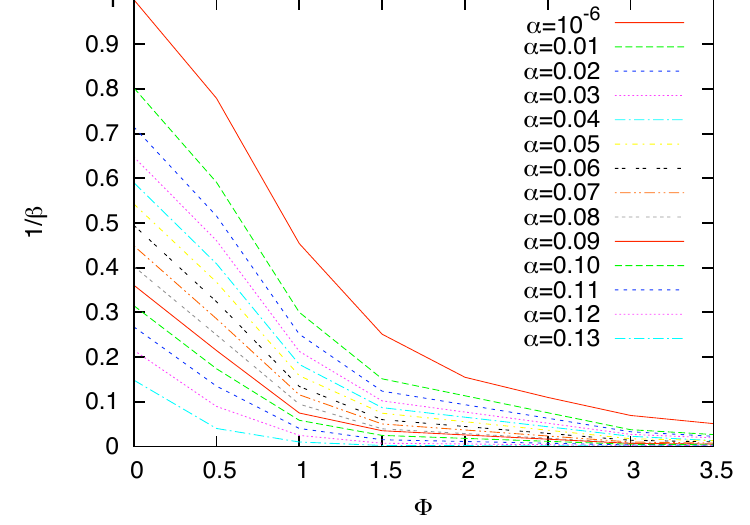}
\caption{Left panels: Phase diagrams given by the critical surface $\beta_M$ at various $\alpha$ (upper panels) and at various $\Phi$ (lower panel). Right panels: Phase diagrams given by the critical surface $\beta_C$ at various $\alpha$ (upper panels) and at various $\Phi$ (lower panel). As usual, $\alpha$ in the plots stands for $\alpha+\gamma$.}
\label{fig:Diagramma}
\end{figure}
Here we study the RS spin-glass and RS pure state solutions numerically: We insert the pure state ansatz $m_\mu=m(1,0,...,0)$ in the self-consistency RS equation system (\ref{AGS1},\ref{AGS2},\ref{AGS3}), then we eliminate the equation for $r$ substituting it in the formers, so to obtain a new set of equations for $m$, $q$ and the free energy (\ref{AGSfree}):
\begin{eqnarray} \label{RSmq} 
m &=& M(m,q;\beta,\alpha+\gamma,\Phi) = \langle \int d\mu(\eta) \int d\mu(z) \tanh\Big(\beta m + \frac{\beta \sqrt{(\alpha+\gamma) q}}{1 - \beta(1 - q)} z
 + \sqrt{\beta}\Phi\eta  \Big)\rangle_{\xi}, \\ \label{RSqm}
q &=& Q(m,q;\beta,\alpha+\gamma,\Phi) = \langle \int d\mu(\eta) \int d\mu(z) \tanh^2\Big(\beta m + \frac{\beta \sqrt{(\alpha+\gamma) q}}{1 - \beta(1 - q)} z
 + \sqrt{\beta}\Phi\eta \Big)\rangle_{\xi},
\end{eqnarray}
\begin{eqnarray} \label{RSfreeenergy}
f(m,q,r) &=& -\frac{\log 2}{\beta}+\frac{\alpha+\gamma}{2}[1+\beta r
(1-q)] + \frac12 \beta \sum_{\mu}m_{\mu}^2 + \\ \nonumber &+&
\frac{\alpha+\gamma}{2\beta}\Big[
\log[1-\beta(1-q)] -\frac{\beta q }{1-\beta(1-q)} \Big] \\
&-& \frac{1}{\beta}\langle \int d\mu(\eta) \int d\mu(z) \log 2
\cosh \Big( \beta m +  \frac{\beta \sqrt{(\alpha+\gamma) q}}{1 - \beta(1 - q)} z
 + \sqrt{\beta}\Phi\eta \Big)\rangle_{\xi}. \nonumber
\end{eqnarray}
We have used the software Wolfram Mathematica $7.0$ to compute numerical solutions of eqs. (\ref{RSmq},\ref{RSqm}) and calculate the free energy (\ref{RSfreeenergy}) of these solutions: To speed up the evaluation, we noticed that the integrand of the RS self-consistency equations is a product of a hyperbolic tangent and two Gaussians. The hyperbolic tangent is always bounded by one and the Gaussians go quickly to zero. As we have fixed the precision of the integration in the software, and therefore the zero, to $10^{-10}$, we decided to fix the extreme of the $z$ and $\eta$ integration to $-5$ and $5$ as  $e^{-25} \sim 10^{-11}$.
\newline
In Figure \ref{intersezioni} the solutions of  $m-M(m,q,1/\beta,\alpha,\Phi)=0$ and $q-Q(m,q,1/\beta,\alpha,\Phi)=0$ for fixed $\alpha=0.01$ and $\Phi=0.5$ at decreasing noise level are shown:
Above a certain level of noise the two lines representing the solutions in the plane $(m,q)$ do not intersect at $m>0$; they cross each other only in a spin-glass state point, $m=0$ and $q>0$. Of course the line $m=0$ is always a solution of $m=M(m,q,\beta,\alpha+\gamma,\Phi)$.
\newline
For every fixed $\Phi$ and $\alpha+\gamma$ there is a noise threshold at which these two lines are tangent, so a pure state solution $m>0$ appears beyond the spin-glass state: We computed all the corresponding free-energies and verified that, for every fixed $\alpha+\gamma$ and $\Phi$, the free energy for the spin-glass state is lower than free energy of the pure RS state until noise is further lowered, so the pure state that appears among candidate solutions is not immediately stable, that is, we are crossing the region of the "spurious states".
At lower levels of noise,  two pure state solutions bifurcate from the former point, both with strictly positive magnetization, $m_1>m_2>0$ and $q_1>q_2>0$. The second solution $(m_2,q_2)$, the one with lower magnetization, has always higher free energy than the first one or the respective spin-glass state, so can be rejected in the thermodynamical sense: Only the higher magnetization pure state is relevant and becomes a global minimum as far as the noise is further  lowered.
\newline
Lastly we know that in the limit of $\beta \rightarrow \infty$ we reach always a pure state with $m=1$ and $q=1$: this is verified for each $\alpha +\gamma < 0.138$ and whatever $\Phi$ and can easily be understood by a scaling argument on the Hamiltonian $\beta\mathcal{H}(h;\xi,\Phi)$\footnote{It is straightforward to see that in the $\beta \to \infty$ limit the random field term can always be neglected with respect to the Hopfield terms.}.
In order to depict the two first order critical surfaces, we have repeated calculations of order parameters and of free-energy in different regions of the space ($\Phi$, $\alpha+\gamma$, $\beta$); in particular, we have calculated free energies  of the pure state and of the spin-glass state solutions collected for different values of $\Phi$, $\alpha+\gamma$ and $\beta$, and compared each other to find the lowest: Where they cross we have the onset of the transition from one phase to another; in Fig.~(\ref{bue}), left panel, we show results of the computation of the free energy for $\Phi=0.5$ at various $\alpha$, while the related Mattis magnetization is depicted in the right panel.
\newline
Starting from the low noise limit and decreasing in $\beta$, the Mattis magnetization suddenly disappears in the $\alpha+\gamma,\Phi$ plane, implicitly defining the critical surface $\beta_M(\alpha,\Phi)$, but only for noise further reduced, namely on the critical surface $\beta_C(\alpha+\gamma,\Phi)$, these minima are the global minima of the free energy (the former are dominated by the underlying spin-glass phase, mirroring the spurious land of the neural counterpart) and so can be labeled as pure states.
\newline
We can see these boundaries ($\beta_M(\alpha+\gamma,\Phi)$ and $\beta_C(\alpha+\gamma,\Phi)$) together, calculated for various values of $\Phi$ in Fig.~\ref{fig:Diagramma}, upper panel.
The curve $\beta_M(\alpha+\gamma,\Phi)$ and $\beta_C(\alpha+\gamma,\Phi)$ in the $(\alpha+\gamma, \Phi)$ plane demarcate different phases. The phase diagram is depicted for several choices of $\Phi$.
Finally, the first-order phase diagram for $\beta_M(\alpha+\gamma,\Phi)$ and $\beta_C(\alpha+\gamma,\Phi)$ critical surfaces at various $\Phi$ all together is shown in Fig.~\ref{fig:Diagramma}, lower panel.
 The phase defined by the $\beta_C$ surface is the one under which every pure state is recalled stably by the network given appropriate initial conditions \footnote{For appropriate condition we intend a state that has non zero significant overlap (greater than $1/\sqrt(H)$ in the finite network) with some stored pattern.}.


\end{document}